\documentclass[12pt]{iopart}

\usepackage{iopams} 
\usepackage{graphicx}
\usepackage{caption}
\usepackage{subcaption}
\usepackage{xcolor}
\usepackage{comment}

\begin{document}

\title[Comparison of MHD and gyrokinetic simulations]{Comparison of MHD and gyrokinetic simulations of linear instabilities at the $q=1$ surface}

\author{F. N. Antlitz$^{1,2}$, X. Wang$^{1}$, M. H\"olzl$^{1}$, G.~T.~A.~Huijsmans$^{3}$, H.~Zhang$^{1}$, J.~Puchmayr$^{1}$, Ph.~Lauber$^{1}$, T.~Hayward-Schneider$^{1}$, B. F.~McMillan$^{4}$, A.~Mishchenko$^{5}$, E.~Poli$^{1}$, Z.X.~Lu$ ^{1}$, JOREK team$^{6}$}

\address{$^1$Max Planck Institut f\"ur Plasmaphysik, Boltzmanstrasse 2,
	D-85748 Garching, Germany}
\address{$^2$Technische Universit\"at M\"unchen, TUM School of Natural Sciences, James-Franck-Straße 1, D-85748 Garching, Germany}
\address{$^3$CEA, IRFM, F-13108 Saint-Paul-lez-Durance, France}
\address{$^4$University of Warwick, Coventry CV4 7AL, United Kingdom}
\address{$^5$Max Planck Institut f\"ur Plasmaphysik, Wendelsteinstraße 1, D-17491 Greifswald, Germany}
\address{$^6$ see the author list of Ref.~\cite{Hoelzl_2024}}
\ead{felix.antlitz@ipp.mpg.de}
\vspace{10pt}
\begin{indented}
\item[]June 2025
\end{indented}

\begin{abstract}
Accurate modeling of core instabilities in tokamak plasmas is essential to understand the underlying physical mechanisms and their impact on plasma confinement. The ideal stability of the internal kink mode and the $m=1$ collisionless tearing mode are analyzed numerically both with gyrokinetic and MHD codes. We compare the different models implemented in the codes and show that the gyrokinetic equations without collisions inherently contain the ideal MHD limit. The simulation results show that the stability of the internal kink mode strongly depends on the choice of several setup parameters like the inclusion of parallel magnetic field fluctuations, the tokamak aspect ratio, the drift- or gyrokinetic treatment of the ions and the electron mass.
Furthermore, we demonstrate the stabilization of the instabilities by diamagnetic effects. 
Our results indicate that gyrokinetic and MHD models can be reconciled in the description of the internal kink mode by careful consideration of the simulation setup and model assumptions, but instabilities like the collisionless tearing mode require a more advanced treatment beyond MHD.

\end{abstract}

%
%
%
%
\section{Introduction}

If the safety factor $q(r)$ falls below unity inside a reference position $r = r_\mathrm{s}$ in a tokamak, linear instabilities in the plasma core with toroidal and poloidal mode numbers $m/n=1/1$ such as the internal kink mode can be excited \cite{Shafranov_1970,Rosenbluth_1973}. These modes can impact the performance of present and future burning plasma tokamak experiments. Both phenomena, so called sawtooth oscillations \cite{vonGoeler_1974} and fishbone bursts \cite{McGuire_1983}, which have been observed regularly in various tokamak devices, are closely connected to the internal kink mode and represent a nonlinear manifestation of this instability. 

The kink mode is triggered by parallel currents in the plasma. In cylindrical geometry (i.e. in a screw pinch), the ideal internal kink is always unstable for $q<1$. However, in toroidal geometry, 
the stability analysis requires increased caution as the toroidal corrections also enter at the lowest non-vanishing order in the inverse aspect ratio as contribution to the perturbed potential energy for $n=1$ \cite{Bussac_1975,Freidberg_2014}. It has been found in \cite{Bussac_1975} that there is a critical value for the poloidal plasma $\beta$ defined by
\begin{equation}
\label{eq:beta_poloidal}
    \beta_\mathrm{p}(r_\mathrm{s}) = -\frac{2 \mu_0}{B_m^2(r_\mathrm{s})}\int_0^{r_\mathrm{s}} \mathrm{d}r \ \left(\frac{r}{r_\mathrm{s}}\right)^2 \frac{\mathrm{d} p}{\mathrm{d} r} = \frac{2 \mu_0}{B_m^2(r_\mathrm{s})} \left[ \left<p\right>_{r_\mathrm{s}} - p(r_\mathrm{s})  \right]
\end{equation}
that needs to be overcome to destabilize the kink mode. Here, $B_m$ is the poloidal magnetic field component, $\mathrm{d}p/\mathrm{d}r$ is the radial pressure gradient and $\left<p\right>_{r_\mathrm{s}}$ is the volume averaged pressure within the $q=1$ surface. For the calculation in \cite{Bussac_1975}, the threshold has been found to be $\beta_\mathrm{p, crit} = \sqrt{13}/12$ for a large aspect ratio tokamak with circular cross section and parabolic current density profile. In the limit $\beta_\mathrm{p} \rightarrow 0$, which represents the case of a plasma with flat pressure profile, the internal kink is ideal MHD stable in toroidal geometry. For this reason, the kink mode is often referred to as both, current- and pressure-driven.

Taking into account finite electrical resistivity leads to deviations from the ideal MHD stability theory. It has been shown that resistivity can increase the growth rate of the kink mode \cite{Coppi_1976}. Furthermore, the resistive kink can be unstable even if the ideal kink is stable. Sometimes the instability is referred to as the ``reconnecting mode'' in this case \cite{Ara_1978}. In this work, however, we will limit the focus on ideal cases with zero resistivity.
But even in the absence of electrical resistivity, magnetic field line reconnection remains possible when considering finite electron inertia. Thus, also in a collisionless plasma, an $m=1/n=1$ instability can be found that is caused by the electron inertia and is appropriately termed ``collisionless tearing mode'' \cite{Drake_1977,Porcelli_1991b}. 

Energetic ions also play an important role in the context of internal kink modes. Experimentally, an increase of the duration between sawtooth crashes with little MHD activity has been observed in discharges with additional ICRH or NBI heating \cite{Campbell_1988}. This behavior was explained theoretically as a consequence of the stabilization of the kink mode by energetic particles (produced by the auxiliary heating in these scenarios) \cite{White_1988,Porcelli_1991a}. The stabilization mechanism is based on the conservation of the third adiabatic invariant.
Adversely, the sawtooth-free period is usually followed by a sawtooth crash with much larger amplitude that can lead to a prompt loss of a significant fraction of the fast particles and trigger other MHD events like neoclassical tearing modes and edge localized modes. These so-called ``giant sawteeth'' are a major concern for future fusion devices. 

On the other hand, if the energetic ion $\beta$ exceeds a critical value, the kink mode is no longer suppressed but a new type of instability, referred to as the fishbone mode is excited. Depending on the properties of the energetic ion population, different branches of the fishbone instability exist \cite{Chen_1984,Coppi_1986,Zonca_2009}. Fishbones can lead to the redistribution and loss of fast ions and thereby limit the efficiency of heating systems, but may be also beneficial for tokamak operation as they are present in discharges with improved confinement \cite{Gruber_1999a,Gruber_1999b,Guenter_2001,Deng_2022,Gao_2018}. An active area of research is the possibility of the generation of sheared flows and the creation of transport barriers by fishbones \cite{Brochard_2024}.

Not only fast ions, but also kinetic effects of thermal ions impact the stability of the internal kink mode. In \cite{Antonsen_1993}, it has been found that trapped thermal particles can substantially stabilize or destabilize the internal kink mode depending on the ion to electron temperature ratio. The synergistic effect of kinetic thermal and fast ions has also been investigated in numerical simulations \cite{Miao_2021}. Inside a thin region near the rational surface $r_\mathrm{s}$, known as the ``inertial layer'', plasma inertia must be retained. This layer can be very narrow, and the fluid description may break down, necessitating a kinetic treatment to accurately capture the relevant physical effects.

Numerous models exist that have been used to address the stability of the kink mode in numerical simulations ranging from pure single or two fluid models \cite{Yu_2015} to fully kinetic or gyrokinetic descriptions \cite{Mishchenko_2012,McClenaghan_2014,Brochard_2022}. Hybrid kinetic-MHD codes are commonly used to include kinetic effects of energetic particles \cite{Park_1992,Bogaarts_2022} and extensions are made to treat the thermal ions kinetically, too \cite{Todo_2021}. A recent publication also demonstrates the simulation of MHD modes with a gyrokinetic code in stellarator geometry \cite{Nuehrenberg_2025}.  

In this work, linear ideal internal kink mode simulations with the global electromagnetic gyrokinetic code ORB5 \cite{Lanti_2020} are compared to the extended MHD code JOREK \cite{Hoelzl_2021}. The exact same scenarios are analyzed with both codes, which allows a direct comparison of the very different physics models implemented in the codes. Although JOREK is equipped with a kinetic particle module, only pure fluid calculations are considered: A standard full MHD model (single fluid) and an extended model with two-fluid diamagnetic extensions. Keeping the considered cases as simple as possible, we focus on the physics mechanisms that play an important role for the $1/1$ instability and identify the relevant effects necessary to accurately capture the instabilities. 

The rest of this paper is structured as follows. In Section~\ref{sec:models}, the different gyrokinetic and MHD models considered are introduced. Subsequently, it is shown that the gyrokinetic equations can be reduced to the MHD limit. In Section~\ref{sec:results}, the simulation results for different 1/1 instabilities are presented, compared and discussed. A conclusion and outlook on future work is given in Section~\ref{sec:conclusion}.






\section{MHD and gyrokinetic models}
\label{sec:models}

\subsection{Gyrokinetic models}

\subsubsection{The ORB5 model.}

ORB5 is a global electromagnetic gyrokinetic initial value code that solves the 5D Vlasov-Maxwell system with multiple species \cite{Lanti_2020}. The field variables are given by the electrostatic potential $\Phi$, and the parallel component of the perturbed magnetic vector potential $A_\parallel$, where the total magnetic field is given by $\mathbf{B} = \mathbf{B}_0 + \mathbf{b}_0\times\nabla A_\parallel$ with the equilibrium magnetic field $\mathbf{B}_0$ and its unit vector $\mathbf{b}_0$. Note that parallel magnetic field fluctuations $\delta B_\parallel$ are not included in the model per se, but can be approximately accounted for via an extension discussed later on. The field equations in the mixed-variable formulation \cite{Mishchenko_2019}, where $A_\parallel = A_\parallel^\mathrm{(s)} + A_\parallel^\mathrm{(h)}$ is split into a symplectic and Hamiltonian part are given by
\begin{equation}
\label{eq:gyrokinetic_Poisson}
    -\nabla\cdot \left[  n_{\mathrm{i}0} \frac{m_\mathrm{i}}{B_0^2} \nabla_\perp \Phi\right] = \sum_{s=\mathrm{i,e}} q_s n_{s1}
\end{equation}
\begin{equation}
	\frac{\partial}{\partial t} A_{\parallel}^{(\mathrm{s})} + \mathbf{b}_0 \cdot \nabla \Phi = 0
\end{equation}
\begin{equation}
    \left[ \sum_{s=\mathrm{i,e}} \mu_0 \frac{q_s^2 n_{s0}}{m_s}  - \nabla_\perp^2 \right] A_{\parallel}^{(\mathrm{h})} = \mu_0 \sum_{s=\mathrm{i,e}} j_{s1\parallel} + \nabla_\perp^2 A_{\parallel}^{(\mathrm{s})}
\end{equation}
$n_s$, $j_s$, $m_s$, $q_s$ are the particle density, current density, mass and charge for species $s$, respectively. Symbols with index 0 and 1 represent equilibrium and perturbed quantities, respectively. ORB5 uses the Coulomb gauge. The distribution function is split into background and perturbed part for each species, too, $f_s = f_{s0} + f_{s1}$. It is assumed that the background part remains constant over time and only the perturbed part is solved ($\delta f$ scheme). Its time evolution is given by
\begin{equation}
\frac{\partial  f_{s1}}{\partial t}+\left.\dot{\boldsymbol{R}} \cdot \frac{\partial f_{s1}}{\partial \boldsymbol{R}}\right|_{v_{\|}}+\dot{v}_{\|} \frac{\partial f_{s1}}{\partial v_{\|}}=-\left.\dot{\boldsymbol{R}}^{(1)} \cdot \frac{\partial f_{0 s}}{\partial \boldsymbol{R}}\right|_{\varepsilon}-\dot{\varepsilon}^{(1)} \frac{\partial f_{0 s}}{\partial \varepsilon}
\end{equation}
with 
\begin{equation}
    \dot{\boldsymbol{R}} = \dot{\boldsymbol{R}}^{(0)} + \dot{\boldsymbol{R}}^{(1)}, \qquad \dot{v}_\parallel = \dot{v}_\parallel^{(0)}  + \dot{v}_\parallel^{(1)} 
\end{equation}
\begin{equation}    
\dot{\boldsymbol{R}}^{(0)} = v_\parallel \mathbf{b}^* + \frac{1}{q_s B_\parallel^*} \mathbf{b}_0\times \mu \nabla B_0
\end{equation}
\begin{equation}
    \dot{v}_\parallel^{(0)} = -\frac{\mu}{m_s} \nabla B_0 \cdot  \mathbf{b}^*  
\end{equation}

\begin{equation}
\dot{\boldsymbol{R}}^{(1)}=  \frac{\mathbf{b}_0}{B_{\|}^*} \times \nabla\left\langle\Phi-v_{\|} A_{\|}^{(\mathrm{s})}-v_{\|} A_{\|}^{(\mathrm{h})}\right\rangle-\frac{q_s}{m_s}\left\langle A_{\|}^{(\mathrm{h})}\right\rangle \mathbf{b}^*     
\end{equation}
\begin{equation}
\dot{v}_{\|}^{(1)}=  -\frac{q_s}{m_s}\left[\mathbf{b}^* \cdot \nabla\left\langle\Phi-v_{\|} A_{\|}^{(\mathrm{h})}\right\rangle+\frac{\partial}{\partial t}\left\langle A_{\|}^{(\mathrm{s})}\right\rangle\right]  -\mu \frac{\mathbf{b}_0 \times \nabla B_0}{m_s B_{\|}^*} \cdot \nabla\left\langle A_{\|}^{(\mathrm{s})}\right\rangle
\end{equation}
\begin{equation}
    \dot{\varepsilon}^{(1)} = {v}_{\|} \dot{v}_{\|}^{(1)} + \frac{\mu}{m_s} \dot{\boldsymbol{R}}^{(1)} \cdot \nabla B_0
\end{equation}
Here, $\left\langle \boldsymbol{\cdot}\right\rangle$ denotes the gyroaverage, $\mu$ is the magnetic moment, $B_\parallel^* = \mathbf{b}_0 \cdot \nabla \times \mathbf{A}^*$, $\mathbf{b}^* = \nabla \times \mathbf{A}^* / B_\parallel^*$ and $\mathbf{A}^* = \mathbf{A}_0 + \left(m_s v_\parallel / q_s\right) \mathbf{b}_0$. In the drift kinetic limit, $v_\parallel$ is the mixed variable $v_\parallel = v_\parallel^\mathrm{(gc)} + q_s / m_s \left\langle A_\parallel^\mathrm{(h)}\right\rangle$, where $v_\parallel^\mathrm{(gc)}$ is the guiding-center parallel velocity. A more general and accurate definition of $v_\|$ can be found in \cite{Mishchenko_2014}.

ORB5 solves the system of equations using a particle-in-cell (PIC) approach. This is done by discretizing $f_{s1}$ by marker particles, which are pushed in time according to the particle equations of motion. Usually, electrons are treated drift-kinetically and ions gyrokinetically. In all the simulations described below, collisions are not accounted for and the initial background magnetic field is calculated self-consistently by the MHD equilibrium condition from the given pressure and safety factor profiles with the CHEASE code \cite{Lutjens_1996}. The background distribution function $f_{s0}$ is taken as a local Maxwellian for all species in all the simulations reported here with a temperature determined by the parameter $l_x = 2 a / \rho_\mathrm{s}$, $a$ being the tokamak minor radius and $\rho_\mathrm{s}$ the ion sound Larmor radius, $\rho_\mathrm{s} = \sqrt{T  m_\mathrm{i}} / \left(q_\mathrm{i} B_\mathrm{ax} \right)$. $B_\mathrm{ax}$ is the magnetic field on axis. For simplicity, in this study ion and electron temperature are assumed to be equal and spatially constant ($T_\mathrm{i} =T_\mathrm{e} = T$).

In order to include the effect of the parallel equilibrium current from the MHD equilibrium, the Maxwellian distribution of the electrons is shifted by $u_0 = j_\mathrm{eq,\parallel} / \left(e n_\mathrm{e}\right)$, i.e. it is assumed that the electrons carry the total current:
\begin{equation}
 f_{\mathrm{e}0} = n_0 \left(\frac{m_\mathrm{e}}{2\pi T_\mathrm{e}} \right)^{3/2} \exp \left[-\frac{m_\mathrm{e} \varepsilon}{T_\mathrm{e}} \right]  \exp \left[-\frac{m_\mathrm{e} u_0 \left( u_0 - 2 v_\parallel\right) }{2 T_\mathrm{e}} \right]
\end{equation}
An important figure of merit is the ratio of the electron thermal velocity and this shift in the Maxwellian distribution. If we approximate $\mu_0 j_\mathrm{eq,\parallel} \approx B_\mathrm{ax}\left(2-\hat{s}\right) / \left( q R_0\right)$, where $\hat{s}$ is the magnetic shear and $R_0$ is the major radius, then
\begin{equation}
    \frac{u_0}{v_\mathrm{th,e}} = \frac{j_\mathrm{eq,\parallel} / \left(e n_\mathrm{e} \right)}{\sqrt{{T_\mathrm{e}}/{m_\mathrm{e}}}}  \approx \sqrt{\frac{m_\mathrm{e}}{m_\mathrm{i}}} \frac{\left(2 - \hat{s}\right)}{q} \frac{2 a}{R_0} \left(l_x \beta_\mathrm{ORB5} \right)^{-1}
\end{equation}
The parameter $\beta_\mathrm{ORB5}$ is defined as 1/2 of the electron plasma $\beta$ 
\begin{equation}
\label{eq:beta_ORB5}
    \beta_\mathrm{ORB5} = \mu_0 \frac{\bar{n}_\mathrm{e} T_\mathrm{e}(s_0) }{B_\mathrm{ax}^2} .
\end{equation}
evaluated with the average electron density $\bar{n}_\mathrm{e}$, magnetic field on axis $B_\mathrm{ax}$ and reference electron temperature $T_\mathrm{e}(s_0)$ (which is equal to the local temperature everywhere in our case).

For the current driven internal kink mode, it is crucial to include the parallel equilibrium current. Therefore the shift in the Maxwellian $u_0$ needs to be well resolved in the simulations. A small plasma $\beta$, small $l_x$, small aspect ratio and large mass ratio $m_\mathrm{e}/m_\mathrm{i}$ is beneficial from a numerical point of view as this increases $u_0 / v_\mathrm{th,e}$ which is typically $\ll 1$.

As mentioned earlier, parallel magnetic field perturbations $\delta B_\parallel$ have not been taken into account in the model so far. However, it is possible to retain the effect of $\delta B_\parallel$ to first order by replacing the drift velocity
\begin{equation}
    \mathbf{v}_\mathrm{d} = \frac{1}{q_s B_0} \mathbf{b}_0\times \left[ m_s v_\parallel^2 \boldsymbol{\kappa} + \mu \nabla B_0 \right] 
\end{equation}
that is used to advance the marker particles in time by  \cite{Tang_1980,Graves_2019,Scott_2024,Waltz_1999}
\begin{equation}
\label{eq:modified_drift_velocity}
    \mathbf{v}_\mathrm{d} = \frac{1}{q_s B_0} \left[ m_s v_\parallel^2  + \mu  B_0 \right] \mathbf{b}_0\times\boldsymbol{\kappa} .
\end{equation}
This approach leverages the perpendicular force balance \cite{Roach_2005}
\begin{equation}
\label{eq:Perpendicular_force_balance}
	\nabla_\perp  \delta p_\perp = -\frac{B_0 \nabla_{\perp}\delta B_\parallel}{\mu_0}
\end{equation}
and makes use of the relation
\begin{equation}
    \mathbf{b}_0\times \nabla B_0 = B_0 \mathbf{b}_0 \times \boldsymbol{\kappa} - 4 \pi \frac{\mathbf{b}_0\times\nabla p_\mathrm{eq}}{B_0}.
\end{equation}

\subsection{MHD models}

\subsubsection{The JOREK model.}
JOREK is a nonlinear extended MHD code with reduced and full MHD formulations and various extensions. Although a kinetic extension has been implemented in the code that enables simulations with kinetic particles for various applications such as energetic particles \cite{Bogaarts_2022}, this paper focuses on the pure fluid models.
The extended visco-resistive full MHD model including diamagnetic terms is given by \cite{Pamela_2020}
\begin{equation}
\label{eq:continuity_equation}
    \frac{\partial \rho}{\partial t} + \nabla \cdot \left[\rho\left(\mathbf{v} + \mathbf{v_\mathrm{*i}} \right)\right]= \nabla\cdot \left[D_\perp \nabla_\perp \rho + D_\parallel \nabla_\parallel \rho\right] 
\end{equation}
\begin{equation}
\label{eq:momentum_equation}
	\rho\left[\frac{\partial }{\partial t} + \left(\mathbf{v}+\mathbf{v_\mathrm{*i}} \right)\cdot\nabla \right]\mathbf{v} = \mathbf{J}\times\mathbf{B} - \nabla p + \mu \nabla^2 \mathbf{v}
\end{equation}
\begin{equation}
\label{eq:pressure_equation}
	\frac{\partial p}{\partial t} = -\mathbf{v} \cdot \nabla p  - \Gamma p \nabla \cdot \mathbf{v} + \nabla\cdot \left[K_\perp \nabla_\perp T + K_\parallel \nabla_\parallel T\right]
\end{equation}
\begin{equation}
\label{eq:induction_equation}
	\frac{\partial \mathbf{A}}{\partial t} = \mathbf{v}\times\mathbf{B} - \eta \left(\mathbf{J} -\mathbf{J}_\mathrm{eq} \right) + \frac{m_\mathrm{i}}{2 e \rho} \nabla_\parallel p
\end{equation}
\begin{equation}
	\mu_0 \mathbf{J} = \nabla\times\mathbf{B} , \quad
	\nabla \cdot \mathbf{B} = 0 
\end{equation}
Note that the Weyl gauge is used in the full MHD model, such that $\partial \mathbf{A} / \partial t = -\mathbf{E}$.
Here, $\rho$ denotes the mass density, $\mathbf{v}$ the MHD velocity, $\mathbf{v}_\mathrm{*i}$ the ion diamagnetic velocity, $\mathbf{J}$ the current density and $p$ the total pressure. The adiabatic index $\Gamma$ is usually set to 5/3 unless stated otherwise. The value of the particle diffusion and heat diffusion coefficients $D_\perp, D_\parallel, K_\perp, K_\parallel$ and the dynamic viscosity $\mu$ and resistivity $\eta$ can be set as input parameters. For comparisons to the ORB5 simulations, which are performed without collisions, all nonideal parameters are set to zero or to a very small value for reasons of numerical stability. In the latter case, scans were carried out to ensure that the simulation results are not influenced by the choice of these parameters. By disabling the diamagnetic terms, which consist of $\mathbf{v_\mathrm{*i}}$ in Equation~(\ref{eq:continuity_equation}) and Equation~(\ref{eq:momentum_equation}), and the parallel pressure gradient term in Equation~(\ref{eq:induction_equation}), the system of equations reduces to a single-fluid MHD model.

Moreover, the equations are further simplified in JOREK's reduced MHD model, which is also used in this work and compared to the full MHD model. It is based on the following ansatz for the magnetic field
\begin{equation}
\label{eq:reduced_MHD_ansatz}
    \mathbf{B} = \frac{F_0}{R} \mathbf{e}_\phi + \frac{1}{R} \nabla \psi \times \mathbf{e}_\phi,
\end{equation}
where $\psi$ is the poloidal magnetic flux.
The toroidal component of the magnetic field is thus constant in time and has a spatial dependence of $1/R$ as $F_0$ is a constant parameter. The formulation of the magnetic field in reduced MHD closely resembles that in ORB5, as the magnetic vector potential can be expressed using only a toroidal component $\mathbf{A} = \psi \nabla \phi$. This form is analogous to using only $A_\parallel$ under the assumption that the parallel direction is predominantly toroidal. 

In the reduced MHD model, a slightly different set of variables is used in the dynamic equation including the velocity stream function $u$, which is proportional to the electrostatic potential $u = \Phi / F_0$, see \cite{Hoelzl_2021} for details. The potentials are not fixed by the choice of a specific gauge, but rather by the ansatz Equation~(\ref{eq:reduced_MHD_ansatz}). In order to compare to the potentials in the ORB5 model, which uses the Coulomb gauge, we compute $\left|\nabla\cdot\mathbf{A} \right| \approx n \left|\psi\right|/R_0^2 $, which is small compared to $B_m$ for a large aspect ratio tokamak and $n=1$. 

Special care needs to be devoted to simulations including the diamagnetic terms in JOREK. By switching them on, finite background flows are building up at the beginning of the simulations as $\nabla \cdot \left( \rho \mathbf{v}_\mathrm{*i} \right) \ne 0$, see Equation~(\ref{eq:continuity_equation}), and therefore the initial condition $\mathbf{v} = 0$ does not allow a static equilibrium. In practice when using the diamagnetic terms, three options exist to run the simulation. The first is to evolve only the $n$=0 component until the flows have established and the system has equilibrated again. Then, the $n$=1 component (and higher harmonics if necessary) is included again and the growing instability is affected by the finite background flows. The second option is to use the initial condition $\mathbf{v} = - \mathbf{v}_\mathrm{*i}$ instead of $\mathbf{v} = 0$. In this way, a finite electric field is already assumed at the beginning of a simulation and it does not need to build up over time. Usually this option is very effective when running the code in the limit $\Gamma=0$ in the equation for the pressure as both, $\partial \rho / \partial t  = 0$ and $\partial p / \partial t  = 0$ initially, see Equation~(\ref{eq:pressure_equation}). The third option is to exclude the evolution of $\rho$ completely in the simulation and keeping $\rho$ constant in time. This is meaningful for linear simulations, as the perturbed density $\rho_1$ does not enter in the linearized equations for the other variables in the case $\Gamma =0$. The density equation is decoupled from the system.

\subsubsection{The CASTOR3D model.} 
As an additional reference for MHD calculations, the linear visco-resistive extended MHD code CASTOR3D \cite{Strumberger_2017} is used, which can be applied to tokamak and stellarator geometry. In the simplest form, the linearized MHD equations solved by CASTOR3D take the form
\begin{equation}
\lambda \rho_1=-\mathbf{v}_1 \cdot \nabla \rho_0-\rho_0 \nabla \cdot \mathbf{v}_1
\end{equation}
\begin{eqnarray}
\lambda \rho_0 \mathbf{v}_1= & -\nabla\left(\rho_0 T_1+\rho_1 T_0\right) / m \\
& +\left[\left(\nabla \times \mathbf{B}_0\right) \times \mathbf{B}_1+(\nabla \times \mathbf{B}_1) \times \mathbf{B}_0\right] / \mu_0
\end{eqnarray}
\begin{equation}
\lambda T_1=-\mathbf{v}_1 \cdot \nabla T_0-(\Gamma-1) T_0 \nabla \cdot \mathbf{v}_1
\end{equation}
\begin{equation}
\lambda \mathbf{B}_1=\nabla \times\left(\mathbf{v}_1 \times \mathbf{B}_0-\eta \nabla \times \mathbf{B}_1 / \mu_0\right)
\end{equation}
These equations represent exactly the linearized version of the base / standard MHD equations in JOREK for the case of vanishing equilibrium velocity $\mathbf{v}_0=0$. A time dependence of the perturbed quantities of the form $\propto\mathrm{e}^{\lambda t}$ was assumed. The exponent $\lambda$ is split into a real and imaginary part $\lambda = \gamma + \mathrm{i}\omega$, where $\gamma$ denotes the growth
rate and $\omega$ the oscillation frequency of the mode \cite{Strumberger_2017}. In CASTOR3D, the equations are formulated as an eigenvalue problem that is solved by the code to obtain the eigenvalues $\lambda$. 

Similar to JOREK, it is also possible to include diamagnetic drift effects via an extended MHD model \cite{Strumberger_2023}. An important detail to mention is the fact, that CASTOR3D defines the variable $\mathbf{v}$ as the total ion velocity (the sum of the $E\times B$, diamagnetic and parallel velocity) rather than the so-called MHD velocity used in JOREK. It is then still assumed that $\mathbf{v}_0 = 0$, which implies that $\mathbf{v}_{E\times B, 0} = -\mathbf{v}_\mathrm{*i, 0}$.

\subsection{Reduction of the gyrokinetic models to the MHD limit}
The MHD vorticity equation can be derived from the ideal MHD model. Here, we linearize the equation of motion (\ref{eq:momentum_equation}) and assume no equilibrium flow $\mathbf{v}_0 = 0$:
\begin{equation}
\label{eq:Linearized_momentum_equation}
    \rho_0 \frac{\partial \mathbf{v}_1}{\partial t} = \mathbf{J}_1 \times \mathbf{B}_0 + \mathbf{J}_0 \times \mathbf{B}_1 - \nabla p_1
\end{equation}
Multiplying this equation by $B_0^{-2} \mathbf{B}_0 \times$ from the left, applying the divergence $\nabla \cdot$, using the linearized Ohm's law $\mathbf{B}_0 \times \mathbf{v}_1 = \mathbf{E}_1$ and the solenoidality of the perturbed current density $\nabla \cdot \mathbf{J}_1 = 0$ yields \cite{Qin_1998}
\begin{eqnarray}
\label{eq:MHD_vorticity}
    \nabla \cdot \frac{\rho_0}{B_0^2}\frac{\partial \mathbf{E}_1}{\partial t} - \nabla \cdot \frac{\mathbf{J}_0 \left[\mathbf{B}_0 \cdot \mathbf{B}_1\right]  - \mathbf{B}_1  \left[ \mathbf{B}_0\cdot\mathbf{J}_0\right] }{B_0^2} & +\left[ \mathbf{B}_0 \cdot \nabla\right]  \frac{ \mathbf{J}_1 \cdot \mathbf{B}_0}{B_0^2}   \nonumber \\
    &+ \nabla p_1 \cdot \nabla \times \frac{\mathbf{B}_0}{B_0^2} = 0 
\end{eqnarray}
This equation can be also recovered from the gyrokinetic models as shown in the following. We start from the conservative form of the gyrokinetic equation 
\begin{eqnarray}
	\frac{\partial}{\partial t} \left[ f_{s1} B_\parallel^* \right] &+ \frac{\partial}{\partial \mathbf{R}} \cdot \left[ \left( \dot{\mathbf{R}}^{(0)} f_{s1} + \dot{\mathbf{R}}^{(1)} f_{s0} \right) B_\parallel^* \right] \nonumber \\
    &+ \frac{\partial}{\partial v_\parallel} \left[\left(\dot{v}_\parallel^{(0)} f_{s1} + \dot{v}_\parallel^{(1)} f_{s0} \right) B_\parallel^* \right] = 0
\end{eqnarray}
and integrate over velocity space and apply the following definitions
\begin{equation}
	n_{s1} (\mathbf{x},t) = \int \mathrm{d}^6 Z B_\parallel^* \, f_{s1} \delta(\mathbf{x} - \mathbf{R})
\end{equation}
\begin{equation}
	j_{s1,\parallel} (\mathbf{x},t) = \int \mathrm{d}^6 Z B_\parallel^* \, f_{s1} q_s v_\parallel \delta(\mathbf{x} - \mathbf{R})
\end{equation}
\begin{equation}
	p_{s1,\parallel} (\mathbf{x},t) = \int \mathrm{d}^6 Z B_\parallel^* \, f_{s1} m_s v_\parallel^2 \delta(\mathbf{x} - \mathbf{R})
\end{equation}
\begin{equation}
	p_{s1,\perp} (\mathbf{x},t) = \int \mathrm{d}^6 Z B_\parallel^* \, f_{s1} \frac{m_s}{2} v_\perp^2 \delta(\mathbf{x} - \mathbf{R})
\end{equation}
Multiplying by the charge, summing over the species and using the quasi-neutrality condition, the equation for the total gyrocenter charge density is 
\begin{eqnarray}
\label{eq:gyrocenter_charge_density}
\frac{\partial }{\partial t} \sum_s q_s n_{s 1} =  \nonumber \\
-\nabla \cdot\left[j_{1\|} \mathbf{b}_0 + p_{1 \perp } \frac{1}{B_0^2} \mathbf{b}_0 \times \nabla B_0+p_{1\|} \frac{1}{B_0}\nabla \times \mathbf{b}_0+j_{\| 0} \delta \mathbf{b}_{\perp}\right]
\end{eqnarray}
with $j_{1 \|} = j_{\mathrm{i}1,\parallel} + j_{\mathrm{e}1,\parallel}$, $p_{1\perp} = p_{\mathrm{i}1,\perp} + p_{\mathrm{e}1,\perp}$, $p_{1\parallel} =p_{\mathrm{i}1,\parallel} + p_{\mathrm{e}1,\parallel}$ and $\delta \mathbf{b}_\perp = \delta \mathbf{B}_\perp / B_0$.

It is important to note that the approximation $1/B_\parallel^* \approx 1/B_0$ was employed in the integrand here. 
As emphasized in \cite{McMillan_2023}, particular care must be taken when relating the parallel velocity moment of the gyrokinetic distribution function to the MHD parallel current in this context. Finite-$\beta$ corrections terms contribute to the parallel current as evaluated from the gyrocenter flux. However, in the limit $\beta\ll 1$ these corrections can be neglected.

Next, we note that the gyrocenter charge density in Equation~(\ref{eq:gyrocenter_charge_density}) is exactly given by Equation (\ref{eq:gyrokinetic_Poisson}) and recognize $\rho_0 = \sum_s m_s n_{s0}$. Using
\begin{eqnarray}
	\nabla p_1 \cdot \left[ \nabla \times \left( \frac{\mathbf{B}_0}{B_0^2} \right) \right] &= \nabla p_1 \cdot \frac{\nabla \times \mathbf{B}_0}{B_0^2} -  \nabla p_1 \cdot \left[ \frac{2}{B_0^3}  \nabla B_0 \times \mathbf{B}_0 \right]  \nonumber \\
    &= \nabla \cdot \left[p_1 \frac{1}{B_0} \nabla \times \mathbf{b}_0 + p_1 \frac{1}{B_0^2} \mathbf{b}_0 \times\nabla B_0\right]
\end{eqnarray}
it becomes evident that the last term in Equation~(\ref{eq:MHD_vorticity}) involving the perturbed pressure is recovered in (\ref{eq:gyrocenter_charge_density}) in the case $p_{\parallel 1} = p_{\perp 1} = p_1$, see also \cite{Deng_2012}. Thus, we arrive at
\begin{equation}
    \frac{\partial }{\partial t} \nabla\cdot \left[\frac{\rho_0}{B_0^2} \nabla_\perp \Phi\right] = \mathbf{B}_0 \cdot \nabla \left( \frac{j_{\parallel 1}}{B_0}\right) + \nabla p_1 \cdot \left[ \nabla \times \left( \frac{\mathbf{B}_0}{B_0^2} \right) \right] + \nabla \cdot  \left(\frac{j_{\parallel 0} \delta \mathbf{B}_\perp}{B_0}\right)
\end{equation}
This is equivalent to Equation (\ref{eq:MHD_vorticity}) in the case $\mathbf{B}_1 = \delta \mathbf{B}_\perp$, i.e. $\mathbf{B}_1 \cdot \mathbf{B}_0 = 0$. Using $\nabla_\perp \Phi =- \mathbf{E}_1$, the term on the left-hand side corresponds to the first term in Equation~(\ref{eq:MHD_vorticity}). The first term on the right-hand side matches the second to last term in Equation~(\ref{eq:MHD_vorticity}). The last term on the right-hand side involves the parallel equilibrium current and the perpendicular component of the perturbed magnetic field and is equivalent to the second term in Equation~(\ref{eq:MHD_vorticity}) for $\delta B_\parallel =0$.

In a consistent manner, the case with $\delta B_\parallel \ne 0$ can be obtained by employing the first order approximation of $\delta B_\parallel$ in ORB5 mentioned earlier and reverting the substitution from Equation~(\ref{eq:Perpendicular_force_balance}). As the term $\mathbf{b}_0\times\nabla B_0 $ is replaced by $\mathbf{b}_0 \times \boldsymbol{\kappa}$, this would just lead to an additional term $\nabla \cdot \left[ p_{\perp 1} \left(\nabla \times \mathbf{B}_0\right)_\perp / B_0^2 \right]$. Using the relation $p_{\perp 1} \approx - B_0 \delta B_\parallel$ reversed, this term becomes
\begin{equation}
    \nabla \cdot \left[ \frac{p_{\perp 1}}{B_0^2} \left(\nabla \times \mathbf{B}_0\right)_\perp \right] \approx \nabla \cdot \left(\frac{\delta B_\parallel}{B_0} \mathbf{j}_{\perp 0} \right),
\end{equation}
which is exactly the contribution from the MHD vorticity equation missing
\begin{equation}
    \nabla \cdot \frac{\mathbf{J}_0 \left[\mathbf{B}_0 \cdot \mathbf{B}_{1\parallel}\right]  - \mathbf{B}_{1\parallel}  \left[ \mathbf{B}_0\cdot\mathbf{J}_0\right] }{B_0^2} = \nabla \cdot \left[\frac{\delta B_\parallel}{B_0} \left(\mathbf{J}_0 - \mathbf{J}_{\parallel 0}\right)\right].
\end{equation}

An alternative derivation demonstrating the recovery of the (reduced) MHD equations from gyrokinetics can be found in \cite{Miyato_2009}.

\subsection{The internal kink mode and its dispersion relation}

In \cite{Coppi_1976}, the growth rate for the internal kink mode in cylindrical geometry ($r,\theta, z$) is not derived from this particular form of the vorticity equation, but similarly, the operator $\mathbf{B}_0 \cdot \nabla \times $ (instead of $\nabla\cdot\left(\mathbf{B}_0 \times\right)$) is applied to Equation~(\ref{eq:Linearized_momentum_equation}). 
This eliminates the perturbed pressure gradient term. Together with the radial component of the induction equation, the well-known equation for the radial displacement $\xi$ can then be concluded from this form of the vorticity equation \cite{Coppi_1976}
\begin{equation}
\label{eq:kink_mode_displacement}
    \frac{\mathrm{d}}{\mathrm{d} r} \left[ r^3 \left(\mu_0 \rho_0 \gamma^2 + F^2\right) \frac{\mathrm{d}\xi}{\mathrm{d} r}\right] - g\xi = 0.
\end{equation}
Here, $\gamma$ is the growth rate, $\xi$ is radial displacement given by $v_{1r} = \gamma\xi$, $F = -\left(B_\theta/r\right) \left(1 - q(r)\right)$, $g=FGr$, $G=F k^2 r^2  + 2 k^2 r^2 B_\theta / r \left(1 + q(r)\right)$.
Solving this equation in an ``outer'' region (away from the rational surface), where plasma inertia is neglected, and in an ``inner'' layer (near the rational surface), where inertia must be retained, and then asymptotically matching the solutions, yields the standard internal kink mode eigenfunction and growth rate in a cylinder. 
The growth rate is given by \cite{Coppi_1976}
\begin{eqnarray}
     \gamma = \frac{\lambda_\mathrm{H}}{\tau_\mathrm{H}}, \quad \lambda_\mathrm{H} = -\frac{\pi}{\left(B_\theta q^\prime(r) r\right)^2_{r=r_\mathrm{s}} } \int_0^{r_\mathrm{s}} \mathrm{d}r \, g, \quad \tau_\mathrm{H} = \frac{r_\mathrm{s}}{\left(v_\mathrm{A\theta}\right)_{r=r_\mathrm{s}}}, \\ 
     \quad v_\mathrm{A\theta}^2 = \frac{B_\theta^2}{\mu_0 \rho} \left(\frac{q^\prime r_\mathrm{s}}{q}\right)^2
\end{eqnarray}

During the derivation in \cite{Coppi_1976} to arrive at Equation~(\ref{eq:kink_mode_displacement}), the incompressibility condition $\nabla\cdot\mathbf{v}_1 = 0$ was used to obtain an equation with only the radial velocity component $v_{1r}$. Similarly, in the derivation of \cite{Ara_1978}, the term $\nabla \left[p_1 + \left(\mathbf{B}_0 \cdot \mathbf{B}_1\right) / \mu_0\right]$ is eliminated using the same argument. This term corresponds to the total (thermal + magnetic) perturbed pressure $p_\mathrm{tot,1} = p_1 + \left(B^2 / \left(2\mu_0\right) \right)_1 \approx p_1 + \mathbf{B}_0\cdot\mathbf{B}_1 / \mu_0 $. This indicates the importance of keeping parallel magnetic field perturbations $\delta B_\parallel$ that largely cancel the perturbed thermal pressure for the kink mode.

In \cite{Ara_1978}, the calculation has been generalized using an extended MHD model and the dispersion relation for the internal kink mode including diamagnetic drift effects has been derived, which introduces factors of $\mathrm{i}\omega_\mathrm{*i/e}$, i.e. adds a real frequency contribution. In the case of a finite equilibrium radial electric field, the dispersion relation must be evaluated in a reference frame rotating with the ${E}\times {B}$ velocity and is given by 
\begin{equation}
\label{eq:dispersion_relation}
    \lambda \left(\lambda - \mathrm{i}\lambda_\mathrm{i}\right) = \lambda_\mathrm{H}^2
\end{equation}
for the ideal internal kink mode. Here, $\lambda = -\mathrm{i}\omega \tau_\mathrm{H}$, $\lambda_\mathrm{i} = -\omega_\mathrm{*i} \tau_\mathrm{H}$   and $\omega$ is the Doppler shifted frequency.
The roots of this dispersion relation can be written as
\begin{equation}
\label{eq:dispersion_relation_omega}
    \omega = \frac{1}{2} \left( \omega_\mathrm{*i} \pm \sqrt{\omega_\mathrm{*i}^2 - 4 \gamma_\mathrm{I}^2} \right) , \quad \gamma_\mathrm{I} = \frac{\lambda_\mathrm{H}}{\tau_\mathrm{H}}.
\end{equation}
Thus, for a highly unstable mode $\gamma_\mathrm{I} > \omega_\mathrm{*i}/2$, the real frequency is half of the ion diamagnetic frequency. In the limit $\gamma_\mathrm{I}\rightarrow 0$, there is a stable solution of the dispersion relation with only a real part $\omega = \omega_\mathrm{*i}$.

Note that from a gyrokinetic perspective, the appearance of the $\omega_*$ terms in the dispersion relation can be attributed to the fact that finite Larmor radius effects are kept in the calculation \cite{Lauber_2018}. This can be seen for example in the gyrokinetic moment equation (Equation (11) in \cite{Lauber_2013}), where the term $\mathbf{b} \times \nabla\left[{\beta_{\mathrm{i} \perp}}/{\left(2 \omega_{\mathrm{ci}}\right)}\right] \cdot \nabla \nabla_{\perp}^2 \Phi$ can be written as $\mathrm{i} \omega_{*\mathrm{i}} / v_\mathrm{A}^2 \nabla_\perp^2 \Phi$ and then combined with the inertia term on the left-hand side, which would lead to the substitution $\omega \rightarrow \omega - \omega_\mathrm{*i}$ \cite{Lauber_2013}. 




\subsection{Simulation setup}

In the following, the results obtained from the numerical experiments with the different gyrokinetic and MHD codes introduced in this section are presented. In order to keep the setup as simple as possible, a tokamak plasma with circular cross section and aspect ratio 10 (minor radius $a=1\,\mathrm{m}$, major radius $R_0 = 10\,\mathrm{m}$ unless stated otherwise) consisting of hydrogen ions and with a magnetic field on the axis of $B_\mathrm{ax}\approx1\,\mathrm{T}$ is considered. Three different scenarios are analyzed. 

\section{Results and discussion}
\label{sec:results}

\subsection{Collisionless tearing mode}

\begin{figure}[htb]
    \centering
    \includegraphics[width=0.5\textwidth]{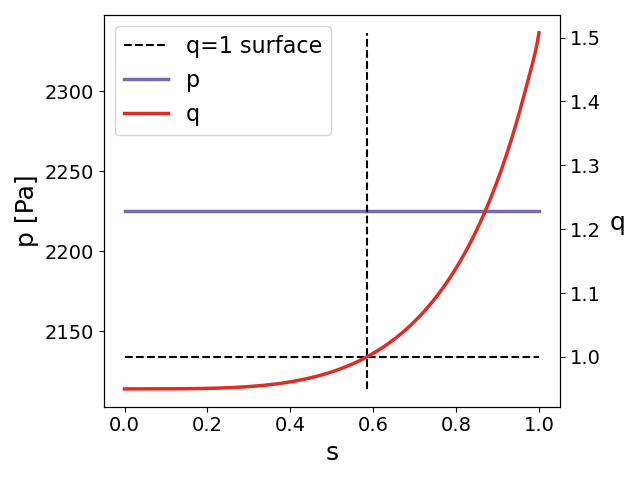}
    \caption{Safety factor and pressure profile for the collisionless tearing mode case. Note that the pressure profile is flat. The radial coordinate used is $s = \sqrt{\psi_\mathrm{N}}$.}
    \label{fig:q_and_p_collisionless_tearing}
\end{figure}

In this section, we consider a case with safety factor and pressure profiles shown in Figure~\ref{fig:q_and_p_collisionless_tearing}. The variable $s=\sqrt{\psi_\mathrm{N}}$, where $\psi_\mathrm{N}$ is the normalized poloidal flux, is used as a radial coordinate. The $q$-profile has a value of 0.95 on axis and increases to 1.5 at $s=1$. The pressure profile is constant over the whole computational domain with $\beta_\mathrm{ORB5} = 0.001385$. The parameter $\beta_\mathrm{ORB5}$ is defined in Equation~(\ref{eq:beta_ORB5}).

Even if there is a $q=1$ surface in the plasma, the internal kink mode is not unstable in toroidal geometry. As explained in the introduction, a threshold for the poloidal $\beta$ defined in Equation~(\ref{eq:beta_poloidal})
needs to be overcome. Hence, a finite pressure gradient is always necessary to destabilize the ideal internal kink mode.

On the other hand, the ``reconnecting'' modes can be driven unstable regardless. In this case, the ``decoupling of the plasma motion from magnetic field lines is required'' to generate magnetic islands and initiate reconnection \cite{Basu_1981}. Even if the plasma is collisionless (and thus the electrical resistivity is zero), the finite electron inertia can lead to this decoupling and enable magnetic reconnection.

Different regimes have been analyzed in the collisionless limit with  kinetic electron~/~fluid ion and kinetic ion~/~fluid electron models \cite{Porcelli_1991b}. In the limit where the ion Larmor radius is smaller than the electron skin depth $\varrho_\mathrm{i} < \delta_\mathrm{e}$, the width of the reconnecting layer is given by $\delta_\mathrm{e}$. Since 
$\delta_\mathrm{e} = \varrho_\mathrm{e} / \sqrt{2\beta_\mathrm{e}}$, this condition can also be written as $\beta_\mathrm{ORB5} < m_\mathrm{e} / m_\mathrm{i}$. In this regime of small plasma $\beta$, the linear growth rate near ideal MHD marginal stability was found to be of the form \cite{Porcelli_1991b,Basu_1981}
\begin{equation}
\label{eq:growth_rate_Basu_Coppi}
    \frac{\gamma}{\omega_\mathrm{A}} \approx \frac{\delta_\mathrm{e}}{r_\mathrm{s}} \propto m_\mathrm{e}^{1/2}
\end{equation}
Here, $\omega_\mathrm{A}$ denotes the Alfv\'en frequency, $\omega_\mathrm{A} = v_\mathrm{A} / L_\mathrm{s}$, $v_\mathrm{A} = B / \sqrt{\mu_0 \rho}$ and $L_\mathrm{s} = R / [r_\mathrm{s} q^\prime (r_\mathrm{s})]$. This scaling has been demonstrated recently in global gyrokinetc simulations of the tearing instability \cite{Jitsuk_2024,Widmer_2024}.

In the limit $\varrho_\mathrm{i} > \delta_\mathrm{e}$, which corresponds to the case $\beta_\mathrm{ORB5} > m_\mathrm{e} / m_\mathrm{i}$, the resistive layer width is determined by $\varrho_\mathrm{i}$ rather than $\delta_\mathrm{e}$. The growth rate given by Equation (\ref{eq:growth_rate_Basu_Coppi}) needs to be modified and becomes 
\begin{equation}
\label{eq:growth_rate_Porcelli}
    \frac{\gamma}{\omega_\mathrm{A}} \approx \left[\frac{2\left(1 + \tau\right)}{\pi} \right]^{1/3}\frac{\delta_\mathrm{e}}{r_\mathrm{s}} \left[\frac{\varrho_\mathrm{i}}{\delta_\mathrm{e}}\right]^{2/3} \propto m_\mathrm{e}^{1/6}
\end{equation}
near ideal MHD marginal stability \cite{Porcelli_1991b}. $\tau$ is the ratio of electron and ion temperature. For $\tau=1$, this growth rate is larger than the one from Equation (\ref{eq:growth_rate_Basu_Coppi}) in the considered regime ($\varrho_\mathrm{i} > \delta_\mathrm{e}$). The instability due to the finite electron inertia found in this regime was called the collisionless $m$=1 tearing mode \cite{Porcelli_1991b}. 

\begin{figure}[htb]
    \centering
    \begin{subfigure}[b]{0.49\textwidth}
         \centering
        \includegraphics[width=\textwidth]{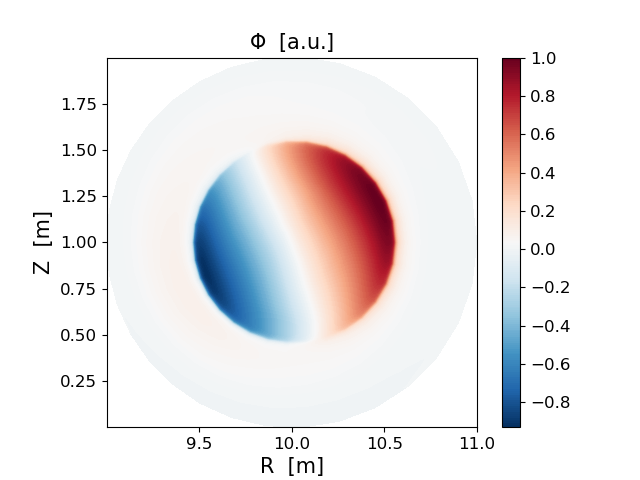}
        \caption{}
        \label{fig:..}
    \end{subfigure}
    \hfill
    \begin{subfigure}[b]{0.49\textwidth}
         \centering
         \includegraphics[width=\textwidth]{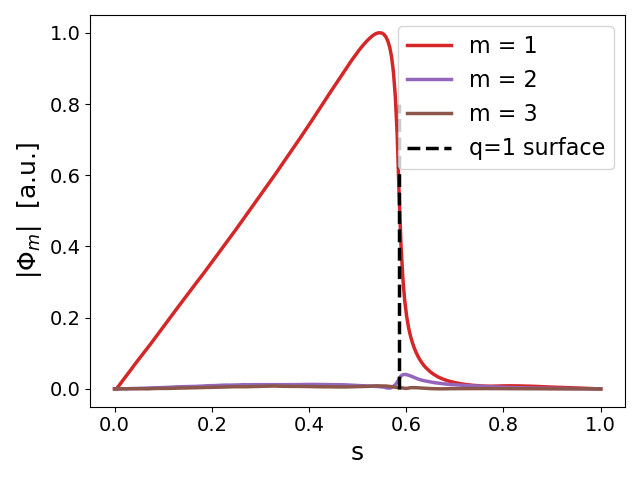}
         \caption{}
         \label{fig:...}
     \end{subfigure}
     \begin{subfigure}[b]{0.49\textwidth}
         \centering
        \includegraphics[width=\textwidth]{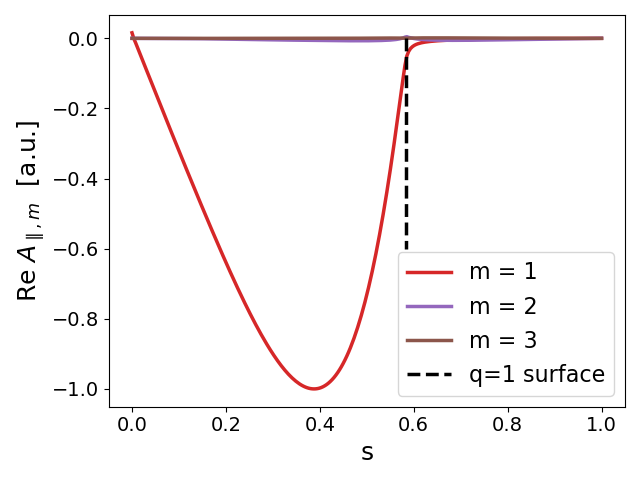}
        \caption{}
        \label{fig:coll_tearing_mode_structure_Apar}
    \end{subfigure}
        \caption{(a) Poloidal and (b) radial mode structure in the electrostatic potential $\Phi$ (absolute value) and real part of the magnetic vector potential component $A_\parallel$ (c) obtained with ORB5 for the case with flat pressure profile and $m_\mathrm{e}/m_\mathrm{i} = 0.001$. }
    \label{fig:coll_tearing_mode_structure}
\end{figure}

\begin{figure}[htb]
    \centering
    \begin{subfigure}[b]{0.49\textwidth}
         \centering
        \includegraphics[width=\textwidth]{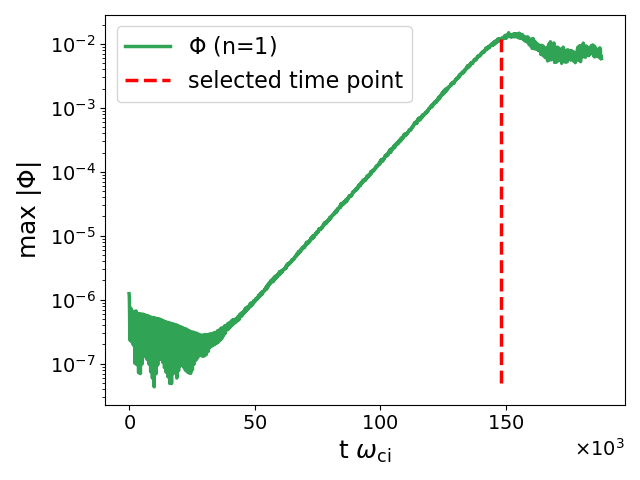}
        \caption{}
        \label{fig:time_trace_coll}
    \end{subfigure}
    \hfill
    \begin{subfigure}[b]{0.4\textwidth}
         \centering
         \includegraphics[width=\textwidth]{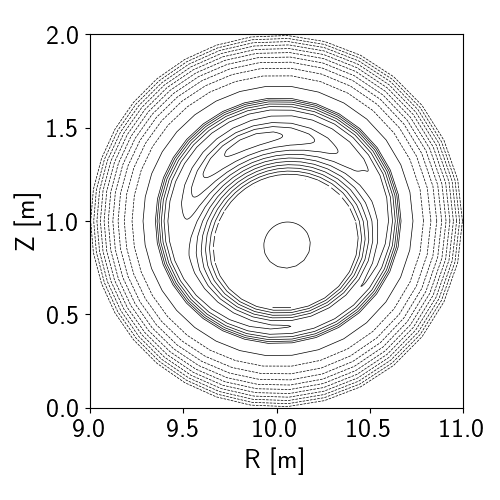}
         \caption{}
         \label{fig:coll_tearing_island}
     \end{subfigure}
        \caption{(a) Time trace of the $n=1$ component of the electrostatic potential $\Phi$ for the ORB5 simulation with $\beta_\mathrm{ORB5} = 0.001385$ and $m_\mathrm{e}/m_\mathrm{i} = 0.005$. The maximum of the absolute value of $\Phi$ (in real space) is taken over the whole computational domain. $\Phi$ is normalized to $T_\mathrm{e} / q_\mathrm{i}$. (b) The helical flux $\Psi_\mathrm{he}$ at $t=148000 \,\omega_\mathrm{ci}^{-1}$ . This time point marks the end of the linear phase when the $n=1$ mode starts to saturate and is represented by the red dashed line in (a). A magnetic island is visible.}
    \label{fig:coll_tearing_mode_time_trace_and_island}
\end{figure}

For the base case considered here, we choose $\beta_\mathrm{ORB5} = 0.001385$ and scan the electron to ion mass ratio $m_\mathrm{e}/m_\mathrm{i}$ in the range 0.0005 to 0.01. Indeed, an $m$=1/$n$=1 dominant instability is found with ORB5. The poloidal and radial mode structure in the electrostatic potential $\Phi$ and magnetic vector potential $A_\parallel$ for the case $m_\mathrm{e}/m_\mathrm{i} = 0.001$ are displayed in Figure~\ref{fig:coll_tearing_mode_structure}. It is important to note that $A_\parallel$ is finite at the $q=1$ surface and the $m=1$ component of $\Phi$ shows a sharp decrease at this position. The temporal evolution of the maximum of the $n=1$ component of $|\Phi|$ is shown in Figure~\ref{fig:time_trace_coll}. At the transition to the nonlinear phase, a magnetic island has grown, which is visible in Figure~\ref{fig:coll_tearing_island}. Here, the contours of the helical flux $\Psi_\mathrm{he} = \psi - \psi_\mathrm{t} / q(r_\mathrm{s}) + \tilde{\psi}$ are displayed, where $\psi$ and $\psi_\mathrm{t}$ are the poloidal and toroidal magnetic flux, $q(r_\mathrm{s}) = m/n = 1$ and $\tilde{\psi}$ is the poloidal flux perturbation.

On the other hand, no instability was found with JOREK for this case. This is consistent as in the single fluid MHD model, the  electrons are assumed to be massless $m_\mathrm{e} \rightarrow 0$. Moreover, the ideal internal kink mode is stable with a flat pressure profile.

\begin{figure}[htb]
    \centering
    \includegraphics[width=0.5\textwidth]{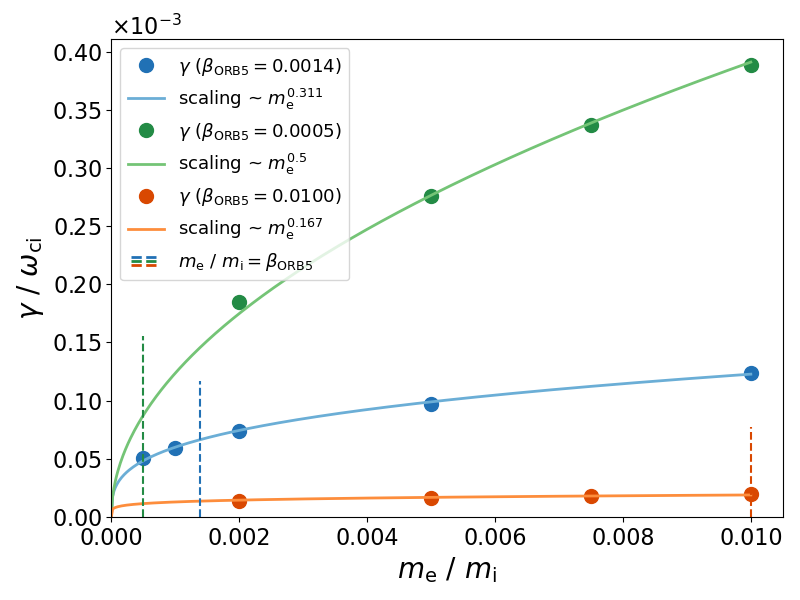}
\caption{The growth rate $\gamma$ of the unstable $m/n=1/1$ mode as a function of the electron-to-ion mass ratio $m_\mathrm{e}/m_\mathrm{i}$ obtained with ORB5 for three different values of $\beta_\mathrm{ORB5}$. The dashed lines represent the value $m_\mathrm{e}/m_\mathrm{i} = \beta_\mathrm{ORB5}$. The green and orange solid lines represent the theoretical scalings $\gamma \propto m_\mathrm{e}^{1/2}$ and $\gamma \propto m_\mathrm{e}^{1/6}$ which have been fitted to the data points. The blue solid line is a fit $\propto m_\mathrm{e}^{\alpha}$, where $\alpha \approx 0.311$ was found as optimal fit parameter.}
    \label{fig:growth_rates_vs_mass_coll_tear}
\end{figure}

Figure~\ref{fig:growth_rates_vs_mass_coll_tear} shows the growth rate of the collisionless tearing mode as a function of the electron to ion mass ratio for three different values of $\beta_\mathrm{ORB5}$. $\beta_\mathrm{ORB5}$ is varied by changing only the density while keeping the temperature constant. The value where $m_\mathrm{e}/m_\mathrm{i} = \beta_\mathrm{ORB5}$ has been marked with a vertical dashed line for each case. For all three data series, the growth rate increases as $m_\mathrm{e}$ is scaled up indicating that the mode is driven by the electron inertia. For the base case ($\beta_\mathrm{ORB5} = 0.001385$), there is a transition between the two regimes mentioned above ($\varrho_\mathrm{i} > \delta_\mathrm{e}$ and $\varrho_\mathrm{i} < \delta_\mathrm{e}$) as the electron mass is increased. We find that the growth rate scales approximately proportional to $m_\mathrm{e}^{0.311}$. This exponent lies in between the two limiting cases from Equations~(\ref{eq:growth_rate_Basu_Coppi}) and (\ref{eq:growth_rate_Porcelli}). The mode becomes much more unstable if $\beta_\mathrm{ORB5}$ is decreased to 0.0005. Now, all data points lie in the regime $\varrho_\mathrm{i} < \delta_\mathrm{e}$. The theoretical scaling of $\gamma \propto m_\mathrm{e}^{1/2}$ is well matched in this case. For $\beta_\mathrm{ORB5}=0.01$, the considered sample points cover the other regime ($\varrho_\mathrm{i} > \delta_\mathrm{e}$). Here, the obtained growth rates from ORB5 scale well with the prediction $m_\mathrm{e}^{1/6}$. 
As can be seen from this figure, the growth rate is significantly reduced by increasing the plasma $\beta$. 
Since it is the goal in the rest of this paper to compare the ideal MHD internal kink between the gyrokinetic and MHD models, it is necessary to be mindful of the influence of the collisionless tearing mode, which becomes critical at low $\beta$ and large electron mass.

\subsection{Stable internal kink mode}

\begin{figure}[htb]
    \centering
    \begin{subfigure}[b]{0.49\textwidth}
        \centering
        \includegraphics[width=\textwidth]{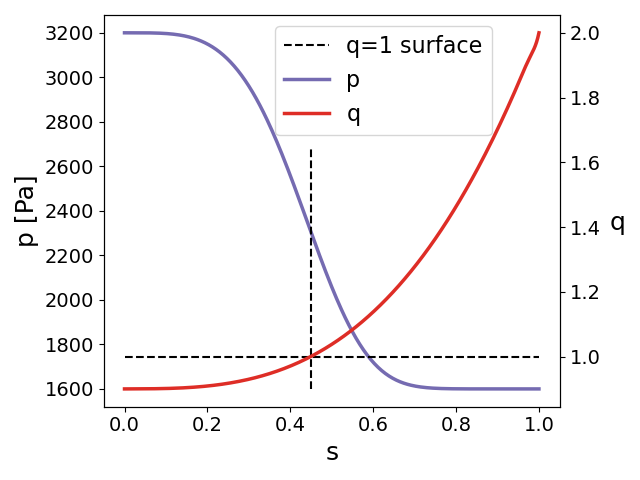}
        \caption{}
        \label{fig:q_and_p_stable_kink}
    \end{subfigure}
    \hfill
    \begin{subfigure}[b]{0.48\textwidth}
        \centering
        \includegraphics[width=\textwidth]{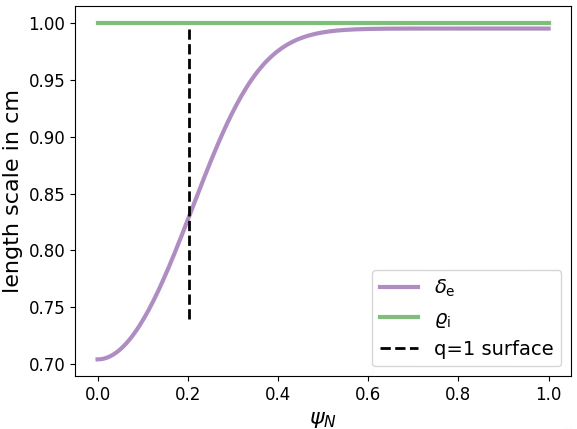}
        \caption{}
        \label{fig:deltae_rhoi_stable_kink}
    \end{subfigure}
    \caption{(a) Safety factor and pressure profile for the stable kink mode case. (b) The electron skin depth $\delta_\mathrm{e}$ and the ion Larmor radius $\varrho_\mathrm{i}$ as a function of the normalized poloidal flux $\psi_\mathrm{N}$ for a scenario where both, the collisionless tearing mode and the internal kink mode are stable. Over the whole domain $\varrho_\mathrm{i} > \delta_\mathrm{e}$.}
    \label{fig:kl}
\end{figure}

Next, we consider an equilibrium with finite pressure gradient and $\delta_\mathrm{e} < \varrho_\mathrm{i}$ over the whole domain. The pressure profile and safety factor profile are shown in Figure~\ref{fig:q_and_p_stable_kink}. The profiles of electron skin depth $\delta_\mathrm{e}$ and ion Larmor radius $\varrho_\mathrm{i}$ are displayed in Figure~\ref{fig:deltae_rhoi_stable_kink}. $\varrho_\mathrm{i}$ is spatially constant because the temperature is set to have a uniform profile. 
When the pressure profile is no longer flat, diamagnetic effects can become important. The dispersion relation for the collisionless tearing mode needs to be modified to take into account corrections by the diamagnetic drift frequency. In \cite{Porcelli_1991b}, it is noted that the collisionless tearing mode is fully stabilized by diamagnetic effects if
\begin{equation}
    1 < \frac{\omega_\mathrm{*e}}{\gamma_\mathrm{0, tear}} = \tau^{1/3} \left(\frac{m_\mathrm{i}}{m_\mathrm{e}}\right)^{1/6} \left(\frac{\beta_\mathrm{e}}{2}\right)^{2/3} \frac{L_\mathrm{s}}{L_n},
\end{equation}
with $L_n = \left| \mathrm{d} \ln n_\mathrm{e} / \mathrm{d} r \right|^{-1} $. $L_\mathrm{s}$ is defined below Equation~(\ref{eq:growth_rate_Basu_Coppi}).

 With $\beta_\mathrm{ORB5} = 0.00115$, $m_\mathrm{e}/m_\mathrm{i} = 0.001$, $L_\mathrm{s} \approx 36 \, \mathrm{m}$ and $L_n \approx 0.40\,\mathrm{m}$, we estimate the ratio of $\omega_\mathrm{*e} / \gamma_\mathrm{0, tear}$, which is the parameter for diamagnetic stabilization of the collisionless tearing mode, to be approximately 3.1. Thus, the collisionless tearing mode should be suppressed.

\begin{figure}[htb]
    \centering
    \begin{subfigure}[b]{0.49\textwidth}
         \centering
         \includegraphics[width=\textwidth]{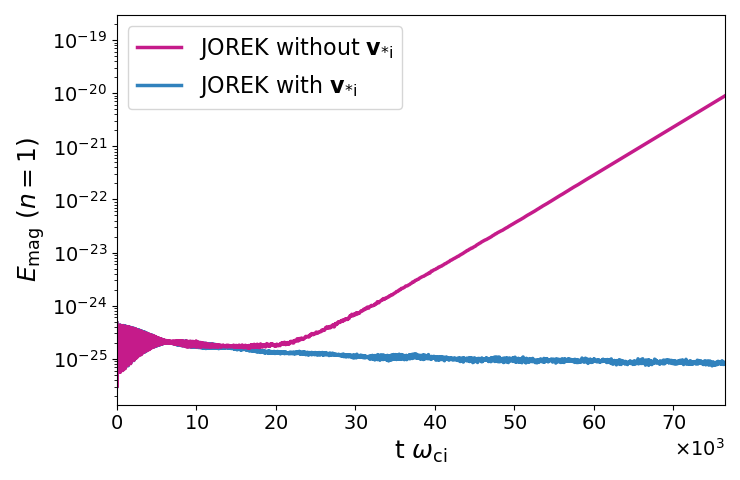}
         \caption{JOREK}
         \label{fig:Jorek_time_trace_stable_kink}
     \end{subfigure}
     \hfill
    \begin{subfigure}[b]{0.49\textwidth}
        \centering
        \includegraphics[width=\textwidth]{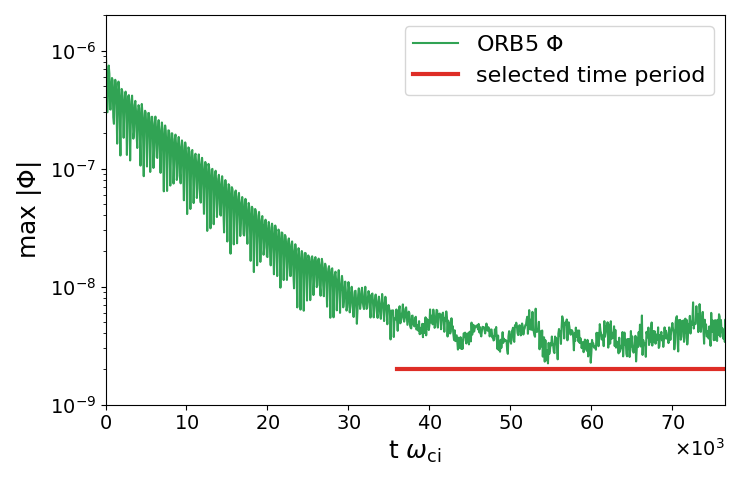}
        \caption{ORB5}
        \label{fig:ORB5_time_trace_stable_kink}
     \end{subfigure}
    \caption{(a) The evolution of the $n=1$ component of the magnetic energy obtained from JOREK simulations with and without diamagnetic terms. (b) Time trace of the electrostatic potential from an ORB5 simulation for the same time period. The maximum of
the absolute value of $\Phi$ (in real space) is taken over the whole computational domain. $\Phi$ is normalized to $T_\mathrm{e} / q_\mathrm{i}$.
The time period, which was used to perform a Fourier analysis is marked by a red line.  }
    \label{fig:time_trace_stable_kink}
\end{figure}

The poloidal plasma $\beta$ defined in Equation~(\ref{eq:beta_poloidal}) is $\beta_\mathrm{p} \approx 0.95$ for this setup. The critical value $\beta_\mathrm{p,crit}$ for the ideal internal kink mode to become unstable can be estimated according to \cite{Porcelli_1996}
\begin{equation}
\label{eq:beta_p,crit}
    \beta_\mathrm{p,crit} = 0.3 \left(1 - \frac{5}{3} \frac{r_\mathrm{s}}{a}\right),
\end{equation}
which is approximately 0.11 for this case. Thus, the ideal internal kink mode is expected to be unstable in the absence of further stabilizing effects. Indeed, the $n=1$ mode is found to be unstable for a JOREK simulation using the full MHD model without diamagnetic drift effects. The temporal evolution of the $n=1$ component of the magnetic energy is shown in Figure~\ref{fig:Jorek_time_trace_stable_kink}.   The growth rate is $\gamma_\mathrm{0, kink} \approx 1.05\cdot 10^{-4} \, \omega_\mathrm{ci}$. 

In comparison, the diamagnetic frequency evaluated at the $q=1$ surface is $\omega_\mathrm{*i} = 6.63\times10^{-4}\,\omega_\mathrm{ci}$. Hence, the ratio of diamagnetic drift frequency and ideal MHD growth rate is $\omega_{*i} / \gamma_\mathrm{0, kink} = 6.3$. For this reason, also the internal kink mode is expected to be fully stabilized by the diamagnetic drift, see Equation~(\ref{eq:dispersion_relation_omega}). In fact, when switching on the diamagnetic terms in the momentum equation in JOREK, no growing mode can be observed in the considered simulation time, see Figure~\ref{fig:Jorek_time_trace_stable_kink} .

\begin{figure}[htb]
    \centering
    \begin{subfigure}[b]{0.49\textwidth}
         \centering
         \includegraphics[width=\textwidth]{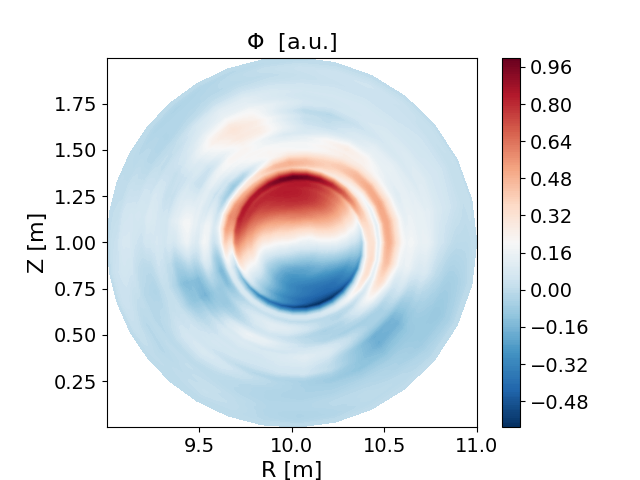}
         \caption{}
         \label{fig:mode_structure_stable_kink}
     \end{subfigure}
     \hfill
     \begin{subfigure}[b]{0.49\textwidth}
         \centering
         \includegraphics[width=\textwidth]{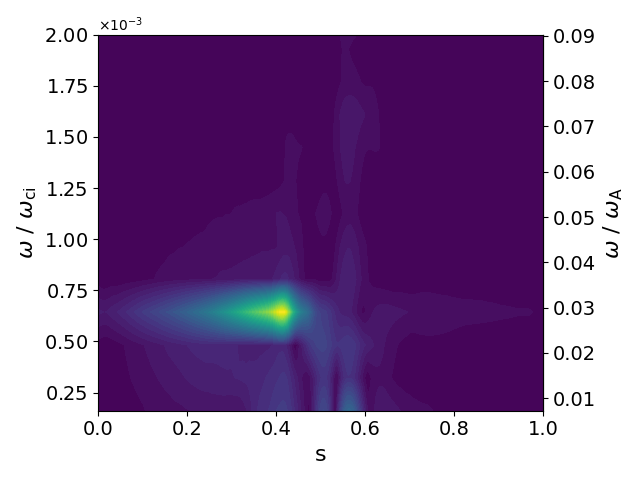}
         \caption{}
         \label{fig:frequency_spectrum_stable_kink}
     \end{subfigure}
    \caption{(a) Snapshot of the poloidal mode structure of the electrostatic potential $\Phi$ at $t=41500\,\omega_\mathrm{ci}^{-1}$ obtained from an ORB5 simulation. (b) Fourier spectrum of $\Phi$ in the second phase of the simulation, marked by the red line in Figure~\ref{fig:time_trace_stable_kink}.}
    \label{fig:mode_structure_spectrum_stable_kink}
\end{figure}

Figure~\ref{fig:ORB5_time_trace_stable_kink} shows the temporal evolution of the $n$=1 component of the electrostatic potential $\Phi$ from an ORB5 simulation in which the diamagnetic effects are automatically included. In a first phase from $t$=0 to $t\approx 36000 \, \omega^{-1}_\mathrm{ci}$, the initial perturbation that is set at the beginning of the simulation decays and reaches a level around which the $n$=1 component of the electrostatic potential oscillates in a second phase from $t\approx 36000 \, \omega^{-1}_\mathrm{ci}$ onward. This is the stable solution of the internal kink mode as $\Phi$ is not growing exponentially in time. The poloidal mode structure is shown in Figure~\ref{fig:mode_structure_stable_kink}. A Fourier analysis is done for this second phase of the simulation. The results, which are shown in Figure~\ref{fig:frequency_spectrum_stable_kink}, reveal that there is one dominant oscillation at $s=0.41$ which corresponds to a location just inside the $q$=1 surface ($s=0.45$). The frequency matches approximately the diamagnetic frequency evaluated at the $q$=1 surface which is $\omega_\mathrm{*i} = 6.63\times10^{-4}\,\omega_\mathrm{ci}$. This is consistent with the dispersion relation in Equation~(\ref{eq:dispersion_relation_omega}) that has the solution $\omega = \omega_\mathrm{*i}$ for $\gamma_\mathrm{I} \approx 0$. 

Additionally, we note that the first phase of the simulation from $t=0$ to $t\approx36000\,\omega_\mathrm{ci}^{-1}$ shows a decaying Alfv\'en eigenmode with much higher frequency, which will not be discussed in further detail here. In order to ensure that the results in the second phase are not affected by the chosen initial condition, another simulation was carried out using an initial perturbation localized only in a narrow region around the $q=1$ surface. In this case, the stable kink mode becomes visible much earlier. The frequency spectrum coincides with that shown in Figure~\ref{fig:frequency_spectrum_stable_kink}.

So far, we considered one scenario where an instability was found with the gyrokinetic model, but not with the MHD model (the collisionless tearing mode) and one scenario where an unstable mode was found using the MHD model (if diamagnetic effects are artificially switched off), but not with the gyrokinetic model (the kink mode stabilized by diamagnetic effects). It is also possible to construct an equilibrium where both, ideal internal kink mode and collisionless tearing mode are stable. This can be done for example by reducing $\beta_\mathrm{p}$ below the critical threshold.
In the following, the more interesting scenario -- where an unstable mode is found with both models in presence of diamagnetic effects -- will be considered in detail.

\subsection{Unstable internal kink}

\begin{figure}[htb]
    \centering
    \includegraphics[width=0.5\textwidth]{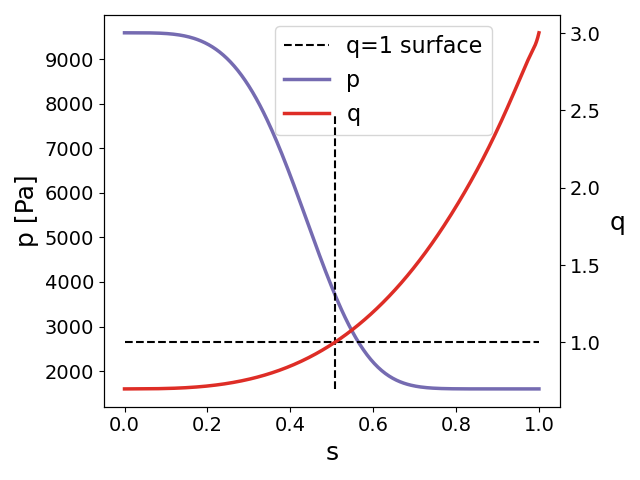}
    \caption{Safety factor and pressure profile for the unstable kink mode case.}
    \label{fig:q_and_p_unstable_kink}
\end{figure}

In this section, we consider cases where the internal kink mode is linearly unstable. Safety factor and pressure profiles of the base case are given in Figure~\ref{fig:q_and_p_unstable_kink}. The $q$ value on axis is 0.7 and increases to 3.0 at $s =1$. The $q=1$ rational surface is located at $s = 0.507$. A low $q$ value on axis was chosen to attain a large growth rate for the internal kink mode and the location of the $q=1$ surface was set to generate a large $\beta_\mathrm{p}$.

\subsubsection{Mode structure, growth rate and frequency.}

\begin{figure}[htbp]
     \centering
     \begin{subfigure}[b]{0.49\textwidth}
         \centering
         \includegraphics[width=\textwidth]{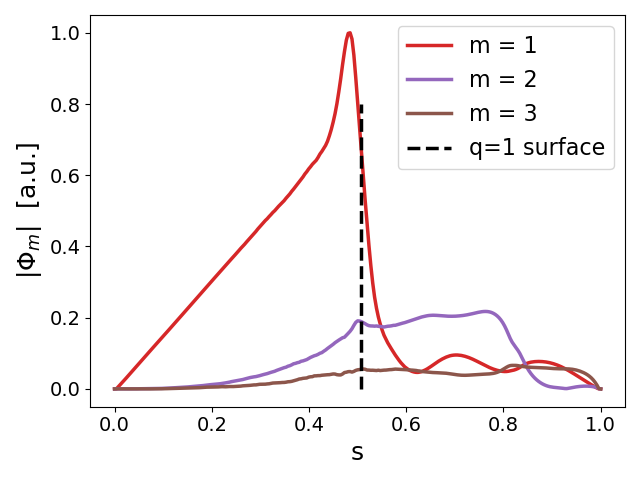}
         \caption{ORB5 without $\delta B_\parallel$ effects}
         \label{fig:Radial_mode_structure_unstable_kink_ORB5_old}
     \end{subfigure}
     \hfill
     \begin{subfigure}[b]{0.49\textwidth}
         \centering
         \includegraphics[width=\textwidth]{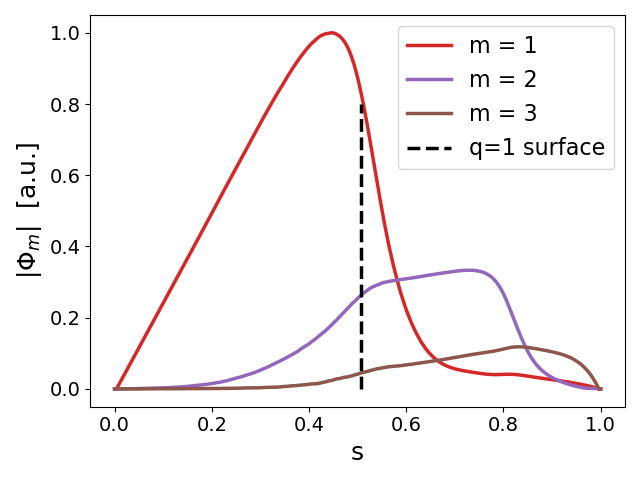}
         \caption{ORB5 with $\delta B_\parallel$ effects}
         \label{fig:Radial_mode_structure_unstable_kink_ORB5}
     \end{subfigure}
     \begin{subfigure}[b]{0.49\textwidth}
         \centering
         \includegraphics[width=\textwidth]{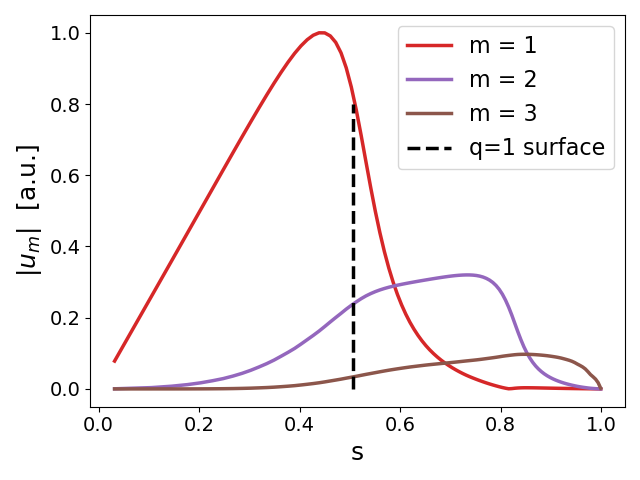}
         \caption{JOREK reduced MHD}
         \label{fig:Radial_mode_structure_unstable_kink_JOREK_redMHD}
     \end{subfigure}
     \hfill
     \begin{subfigure}[b]{0.49\textwidth}
         \centering
         \includegraphics[width=\textwidth]{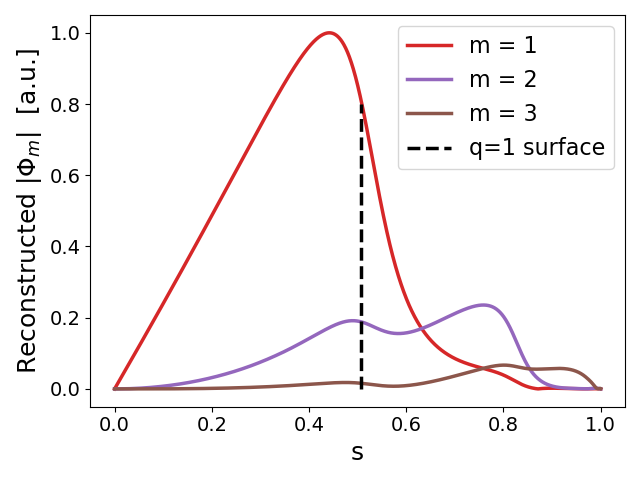}
         \caption{JOREK full MHD}
         \label{fig:Radial_mode_structure_JOREK_fMHD_reconstructedPhi}
     \end{subfigure}
        \caption{Comparison of the radial mode structure for the unstable kink mode case for $\beta_\mathrm{ORB5} = 0.0016$. For ORB5, the absolute value of the electrostatic potential is plotted which is the same quantity as the velocity stream function $u$ in JOREK reduced MHD up to a factor $F_0$. For JOREK full MHD, the electrostatic potential potential is not included in the model, hence it needs to be reconstructed in the Coulomb gauge by solving the Poisson equation~(\ref{eq:Poisson_equation_for_potential_fMHD}). }
        \label{fig:Radial_mode_structure_unstable_kink}
\end{figure}

\begin{figure}[htbp]
\centering
     \begin{subfigure}[b]{0.49\textwidth}
         \centering
         \includegraphics[width=\textwidth]{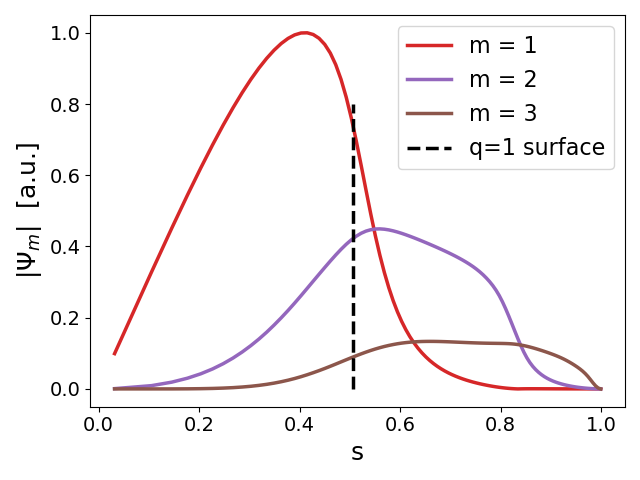}
         \caption{JOREK full MHD}
         \label{fig:Radial_mode_structure_unstable_kink_JOREK_Psi}
     \end{subfigure}
     \hfill
     \begin{subfigure}[b]{0.49\textwidth}
         \centering
         \includegraphics[width=\textwidth]{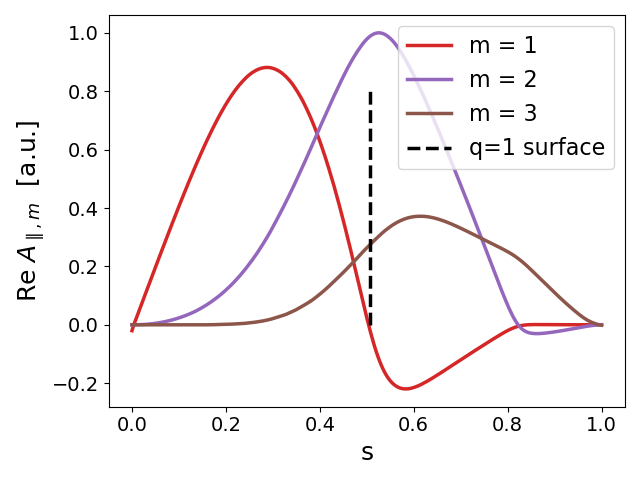}
         \caption{ORB5 with $\delta B_\parallel$ effects}
         \label{fig:Radial_mode_structure_unstable_kin_Apar}
     \end{subfigure}
     \begin{subfigure}[b]{0.49\textwidth}
         \centering
         \includegraphics[width=\textwidth]{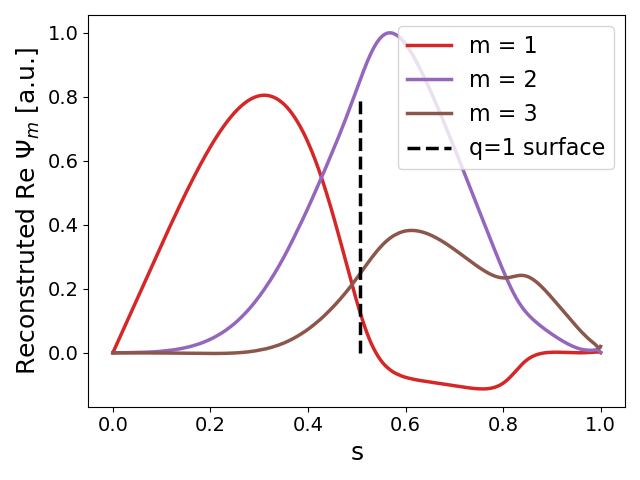}
         \caption{JOREK full MHD}
         \label{fig:Radial_mode_structure_unstable_kink_JOREK_Psireconstruced}
     \end{subfigure}
        \caption{The radial mode structure for the unstable kink mode case for $\beta_\mathrm{ORB5} = 0.0016$. (a) The absolute value of the poloidal magnetic flux $\Psi = R A_\phi$ for the simulation with the JOREK full MHD model, which employs the Weyl gauge. (b) The real part of $A_\parallel$ for the ORB5 simulation including $\delta B_\parallel$ effects. (c) The real part of $\Psi$ reconstructed in the Coulomb gauge from (a).} 
        \label{fig:Radial_mode_structure_unstable_kink2}
\end{figure}

In Figure~\ref{fig:Radial_mode_structure_unstable_kink}, the radial mode structure of the $n$=1 component of the emerging instability is compared between JOREK's reduced MHD, full MHD and ORB5's models with and without $\delta B_\parallel$ effects. The electrostatic potential $\Phi$ is plotted for ORB5, which can be directly compared to the velocity stream function $u$ in reduced MHD. In the full MHD model of JOREK, $\Phi =0$ by choice of the gauge. The poloidal flux $\Psi = R A_\phi$ is shown instead in Figure~\ref{fig:Radial_mode_structure_unstable_kink_JOREK_Psi}. For a direct comparison, we reconstruct the electrostatic potential in the Coulomb gauge starting from the known vector potential $\mathbf{A}$ and solving the Poisson equation
\begin{equation}
\label{eq:Poisson_equation_for_potential_fMHD}
    \nabla^2 \chi_{m,n} \approx \left[ \frac{1}{r} \frac{\partial }{\partial r} \left(r \frac{\partial }{\partial r} \right) - \frac{m^2}{r^2} \right] \chi_{m,n} = f_{m,n} (r)
\end{equation}
for each Fourier component $m/n$, where $f_{m,n}$ are the Fourier components of $\nabla\cdot\mathbf{A}$. The reconstructed electrostatic potential is then given by $\Phi = \partial \chi / \partial t$ and displayed in Figure~\ref{fig:Radial_mode_structure_JOREK_fMHD_reconstructedPhi}.

 The mode structure in Figure~\ref{fig:Radial_mode_structure_unstable_kink_ORB5_old} is obtained without the replacement of the drift velocity to take into account the effect of $\delta B_\parallel$, whereas in Figure~\ref{fig:Radial_mode_structure_unstable_kink_ORB5} the replacement has been made. There are significant differences between these two cases, especially for the $m=1$ component that features a pronounced peak near the $q$=1 surface. The mode structure for the ORB5 simulation with the drift velocity replacement is very similar to the one obtained with both JOREK models. 

In Figure~\ref{fig:Radial_mode_structure_unstable_kin_Apar}, the mode structure in $A_\parallel$ is plotted for the ORB5 simulation with $\delta B_\parallel$ effects. In contrast to the collisionless tearing mode (see Figure~\ref{fig:coll_tearing_mode_structure_Apar}), the different $m$ components of $A_\parallel$ change sign at the corresponding rational surfaces here. Note that $\Psi$ in Figure~\ref{fig:Radial_mode_structure_unstable_kink_JOREK_Psi} does not show this behavior because of the different gauge. However, $R A_\phi$ can be also reconstructed in the Coulomb gauge (Figure~\ref{fig:Radial_mode_structure_unstable_kink_JOREK_Psireconstruced}) and then reveals a similar mode structure compared to Figure~\ref{fig:Radial_mode_structure_unstable_kin_Apar}. This is done using $\mathbf{A}^\mathrm{(Coulomb)} = \mathbf{A}^\mathrm{(Weyl)} - \nabla \chi$.

\begin{figure}[htb]
     \centering
     \begin{subfigure}[b]{0.49\textwidth}
         \centering
         \includegraphics[width=\textwidth]{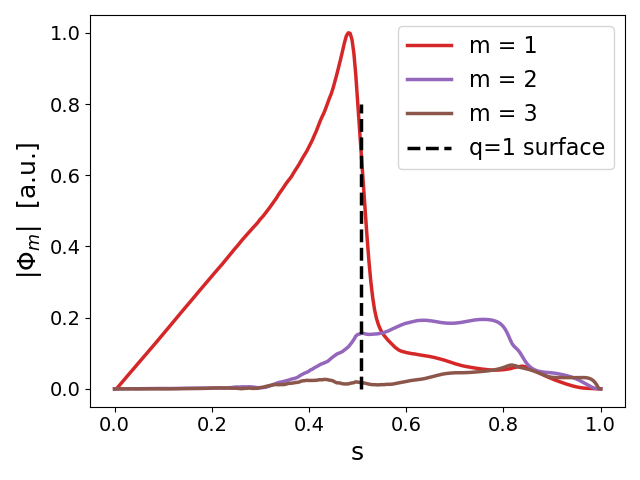}
         \caption{$\beta_\mathrm{ORB5}$ = 0.0012, without $\delta B_\parallel$ effects}
         \label{fig:bla}
     \end{subfigure}
     \hfill
     \begin{subfigure}[b]{0.49\textwidth}
         \centering
         \includegraphics[width=\textwidth]{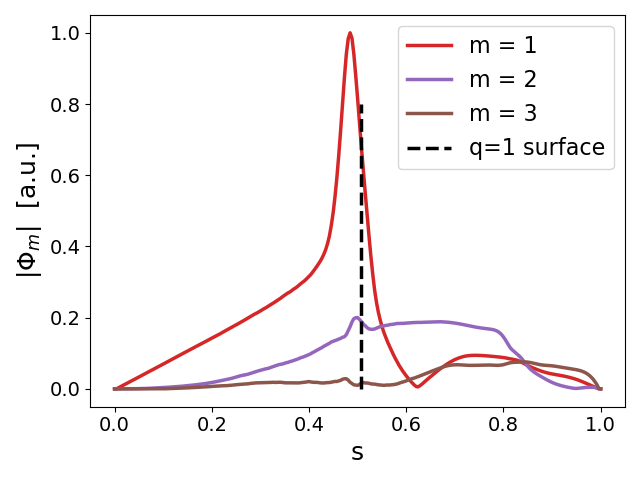}
         \caption{$\beta_\mathrm{ORB5}$ = 0.0020, without $\delta B_\parallel$ effects}
         \label{fig:blabla}
     \end{subfigure}
     \begin{subfigure}[b]{0.49\textwidth}
         \centering
         \includegraphics[width=\textwidth]{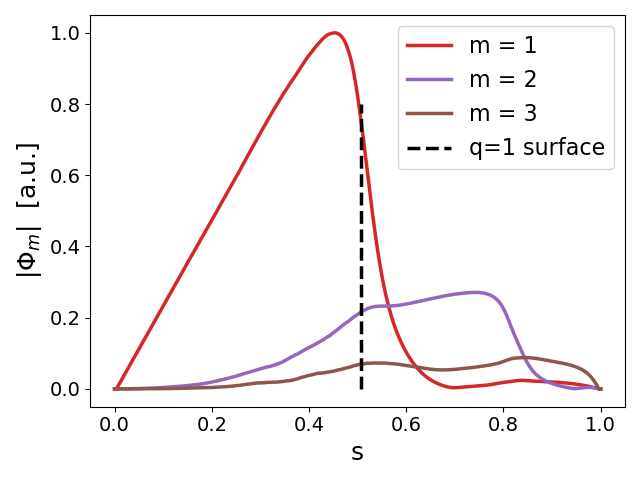}
         \caption{$\beta_\mathrm{ORB5}$ = 0.0012, with $\delta B_\parallel$ effects}
         \label{fig:bla}
     \end{subfigure}
     \hfill
     \begin{subfigure}[b]{0.49\textwidth}
         \centering
         \includegraphics[width=\textwidth]{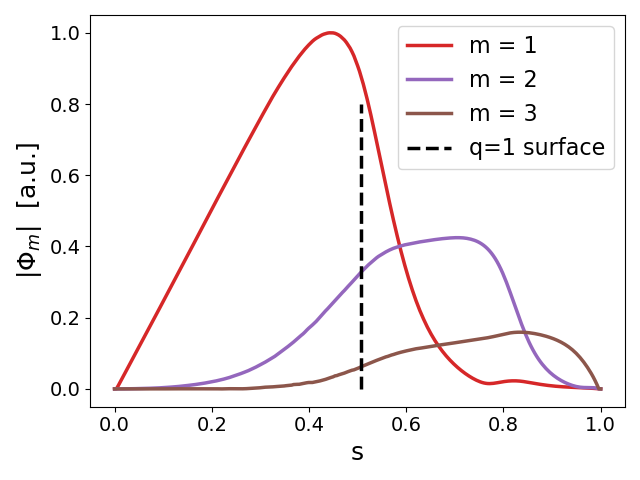}
         \caption{$\beta_\mathrm{ORB5}$ = 0.0020, with $\delta B_\parallel$ effects}
         \label{fig:blabla}
     \end{subfigure}
        \caption{Radial mode structure obtained by ORB5 simulations with and without $\delta B_\parallel$ effects for two values of $\beta_\mathrm{ORB5}$. }
        \label{fig:Radial_mode_structure_unstable_kink_beta_scan}
\end{figure}

A scan in the plasma $\beta$ shows that the differences in ORB5 with and without taking into account $\delta B_\parallel$ become more significant at larger values of $\beta$, see Figure~\ref{fig:Radial_mode_structure_unstable_kink_beta_scan}. $\beta$ is varied by changing the density and keeping the temperature constant. When neglecting $\delta B_\parallel$ effects, the peak near the rational surface becomes more dominant at the higher value of $\beta$, while the mode structure becomes more similar to the MHD results at the lower value of $\beta$. In contrast, a significant change in the mode structure with $\beta$ is not observed in the case when taking $\delta B_\parallel$ into account. At higher $\beta$, the mode structure only becomes broader as poloidal harmonics couple more strongly.

\begin{figure}[htb]
     \centering
     \begin{subfigure}[b]{0.9\textwidth}
         \centering
         \includegraphics[width=\textwidth]{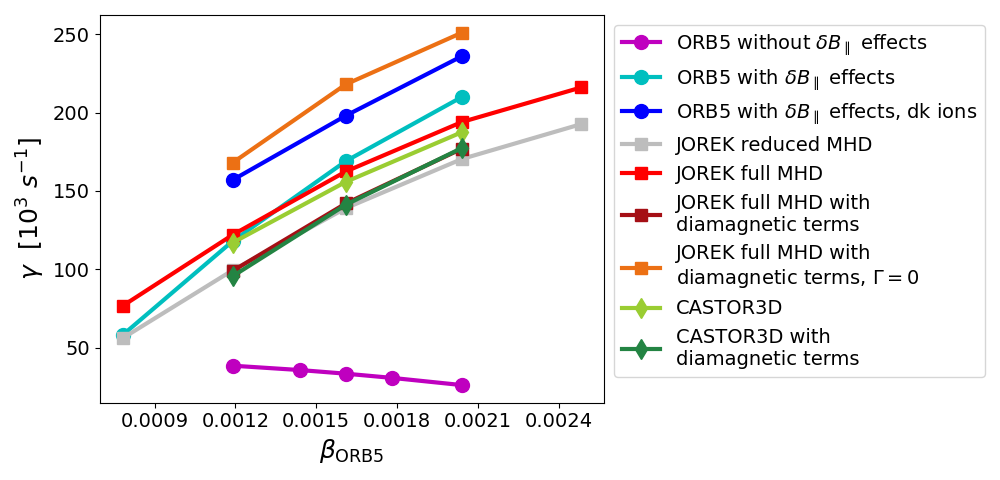}
         \caption{}
         \label{fig:growth_rates_beta_scan}
     \end{subfigure}
     \begin{subfigure}[b]{0.85\textwidth}
         \centering
         \includegraphics[width=\textwidth]{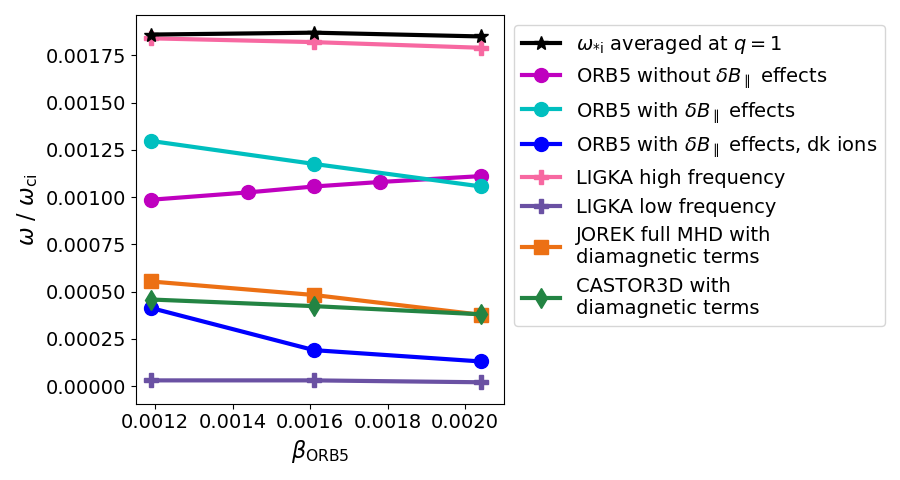}
         \caption{}
         \label{fig:frequencies_beta_scan}
     \end{subfigure}
        \caption{Comparison of the (a) growth rates and (b) real frequencies obtained from the different numerical codes and their models for a scan in the parameter $\beta_\mathrm{ORB5}$.}
        \label{fig:growth_rates_and_frequencies_beta_scan}
\end{figure}

The linear growth rates as a function of the plasma $\beta$ are compared in Figure~\ref{fig:growth_rates_beta_scan} between the different models. The growth rates for the JOREK full MHD simulations are always larger than for the reduced MHD runs when turning off the diamagnetic terms. However, they differ only by approximately 10\% to 25\%  and both show the same tendency to increase with $\beta$. This behavior is expected, as it is well known that the stabilizing term in the MHD energy functional associated with fast magnetosonic waves is reduced in full MHD \cite{Hoelzl_2021}. Yet at low $\beta$, the effect from this term is small and full and reduced MHD lead to similar results, which was already shown in previous simulations with JOREK \cite{Pamela_2020}. Including the diamagnetic terms reduces the growth rate slightly in the simulations with the full MHD model and an adiabatic index of $\Gamma=5/3$. This is behavior expected as well, see Equation~(\ref{eq:dispersion_relation_omega}). When $\Gamma$ is set to zero so that the compression term $\Gamma p \nabla \cdot \mathbf{v}$ is eliminated from the pressure equation, the growth rate considerably increases, even when diamagnetic terms are turned on. 

The growth rates obtained with ORB5 show a strong dependency on how the simulation is set up. By using the standard drift velocity in the equations (without the $\delta B_\parallel$ effect), the growth rate is much smaller compared to all other simulations and also shows the opposite trend with $\beta$. On the other hand, when taking into account the replacement in the drift velocity, the growth rate in ORB5 matches well the reduced and full MHD results in JOREK when using gyrokinetic ions. In the case of drift kinetic (dk) ions, the growth rate becomes even larger, but is still comparable to the MHD results with $\Gamma=5/3$ and still lower than in the case $\Gamma=0$ and follows the same trend. This shows that there is a stabilization of the kink mode when taking into account the ion gyro-average. 

The growth rates calculated from the CASTOR3D code are in good agreement with the JOREK full MHD results. In both cases -- with and without diamagnetic terms -- there is an excellent match between the values for the growth rate. This serves as a benchmark since the physics model in CASTOR3D and JOREK full MHD is the same.

Figure~\ref{fig:frequencies_beta_scan} shows the mode frequency as a function of the plasma $\beta$. Without diamagnetic terms, the internal kink mode in JOREK is always static, which is consistent with linear theory \cite{Ara_1978}. A finite real frequency of the mode can be observed in the simulations with JOREK only when the diamagnetic terms are enabled. Here, we use the option $\mathbf{v}=-\mathbf{v}_\mathrm{*i}$ as an initial condition. This matches the assumption in CASTOR3D that the $E\times B$ and diamagnetic velocity cancel and thus the unperturbed total ion velocity is zero. The obtained frequencies with JOREK and CASTOR3D coincide well and exhibit a slight decreasing trend as $\beta$ increases. Their value is about $\omega_\mathrm{*i} / 4$ evaluated at the $q$=1 surface. In ORB5, the mode frequency is slightly increasing with $\beta$ in the  case where $\delta B_\parallel$ effects are included, while it is decreasing with $\beta$ if they are not. The value of $\omega$ is close to half of the diamagnetic frequency $\omega_\mathrm{*i}$ for gyrokinetic ions. For drift-kinetic ions, the frequency is substantially reduced. This is also consistent with the smaller stabilization seen when neglecting the gyro-average in the drift-kinetic case. 

The present case is also analyzed using the linear gyrokinetic eigenvalue code LIGKA \cite{Lauber_2007}. The code is run using its antenna model to scan a broad frequency range and identify the frequency at which the kink mode is resonantly excited. Two resonance frequencies for each value of $\beta$ are found which are also displayed in Figure~\ref{fig:frequencies_beta_scan}. One resonance is located at $\omega\approx0$ and the other at $\omega\approx\omega_{*i}$. These represent the two stable solutions of the dispersion relation Equation (\ref{eq:dispersion_relation_omega}) in the limit $\gamma_\mathrm{I} = 0$.
The mode structure obtained with LIGKA at these frequencies matches the MHD results well. 
%
The analysis of the strongly unstable mode is not possible with the antenna method -- it requires a different solver (inverse vector iteration). The details, including numerical convergence studies, will be published elsewhere.

\subsubsection{Diamagnetic rotation and comparison to the analytical dispersion relation.}

It can be shown from a gyrokinetic calculation (using LIGKA's model) that the $\omega_*$ terms in the dispersion relation only enter when taking into account finite Larmor radius correction \cite{Lauber_2018, Lauber_2013}. Intuitively, this is also clear by imagining the diamagnetic drift as the net velocity resulting from adding the particles' gyromotion in a plasma with nonzero density or temperature gradient. Thus with only drift kinetic ions, the mode frequency is expected to be close to zero. 

\begin{figure}[htb]
    \centering
    \includegraphics[width=0.5\textwidth]{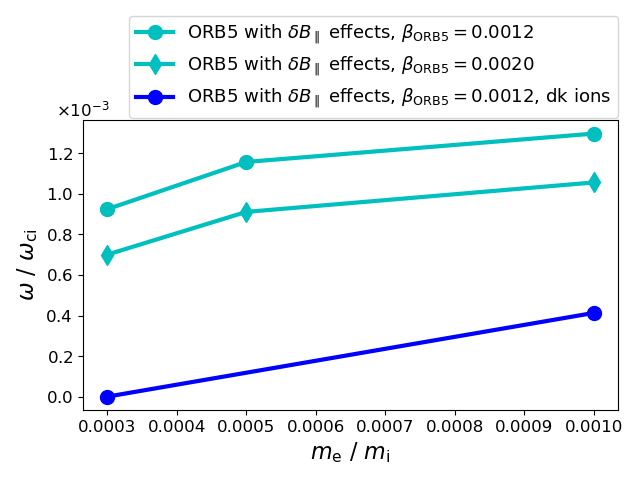}
    \caption{The mode frequency obtained from ORB5 simulations including $\delta B_\parallel$ effects as a function of the electron-to-ion mass ratio. The light blue data points show results from simulations with gyrokinetic ions for two values of $\beta_\mathrm{ORB5}$ while the dark blue data points shows show those with drift kinetic ions. }
    \label{fig:frequencies_vs_e_mass_ORB5}
\end{figure}

In fact, the finite electron mass still shows influence on the mode frequency. The dependence of $\omega$ on the ion-to-electron mass ratio for fixed plasma $\beta$ is displayed in Figure~\ref{fig:frequencies_vs_e_mass_ORB5}. Here, the electron mass has been scaled down to even lower values than the realistic one, which leads to smaller mode frequencies. For the case with drift-kinetic ions, the mode frequency becomes close to zero at $m_\mathrm{e} / m_\mathrm{i} = 0.0003$. The results suggest that the finite electron mass affects the frequency in particular at low $\beta$, which is in line with the observations regarding the collisionless tearing mode. The growth rate is barely affected when reducing the electron mass -- it changes by less than 5\% in the most extreme case -- indicating convergence in this parameter. 

\begin{figure}[htb]
    \centering
    \begin{subfigure}[b]{0.49\textwidth}
        \centering
        \includegraphics[width=\textwidth]{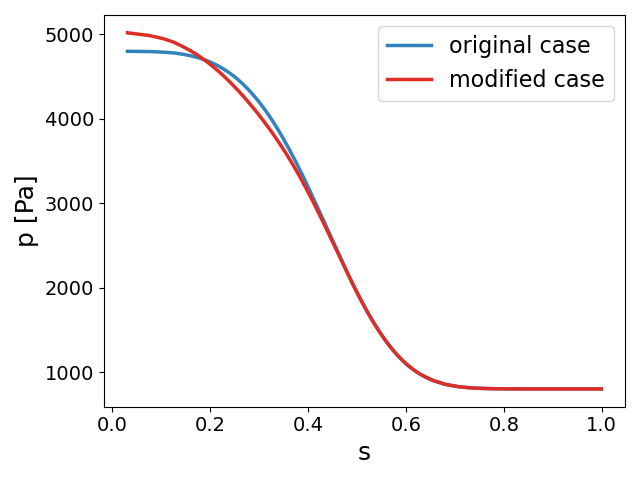}
        \caption{}
         \label{fig:modified_pressure_profile_for_const_omegastar}
     \end{subfigure}
     \hfill
    \begin{subfigure}[b]{0.49\textwidth}
        \centering
        \includegraphics[width=\textwidth]{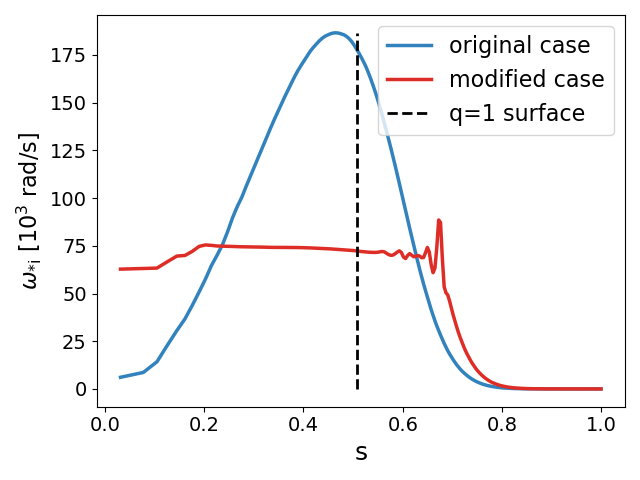}
        \caption{}
        \label{fig:modified_omega_star_profile}
     \end{subfigure}
     \caption{(a) The pressure profile is slightly adjusted in the center to generate a finite gradient near the magnetic axis in order to keep $\omega_\mathrm{*i}$ approximately constant within the $q=1$ surface. (b) Profile of the ion diamagnetic frequency $\omega_\mathrm{*i}$ for the original case and the case with adjusted temperature and density profiles. }
     \label{fig:}
\end{figure}

These findings indicate that the mode frequency obtained in the gyrokinetic simulations approaches the one found with MHD codes in the limit $m_\mathrm{e} \rightarrow 0$. In analytical theory, the mode rotation frequency is expected to be $\omega_\mathrm{*i} / 2$ in extended MHD, see Equation~(\ref{eq:dispersion_relation_omega}). The values obtained in JOREK and CASTOR3D, which are lower than $\omega_\mathrm{*i}/2$ evaluated at $q=1$, can be explained as a consequence of the fact that $\omega_\mathrm{*i}$ is not uniform across the spatial domain. The profile of $\omega_\mathrm{*i}$ as a function of the radial coordinate $s$ is shown in Figure~\ref{fig:modified_omega_star_profile}. Its maximum is located just before the rational surface, but it is approaching zero at the magnetic axis. Hence, the mode frequency is expected to be rather an average of $\omega_\mathrm{*i}(r)/2$ inside the $q=1$ surface.  


To confirm this hypothesis with JOREK, we adjust the density and temperature profiles to achieve a less strongly varying ion diamagnetic frequency. For this purpose, the pressure profile is slightly modified in the center such that a finite pressure gradient exists near the magnetic axis, see Figure~\ref{fig:modified_pressure_profile_for_const_omegastar}. By adapting the ratio of density and temperature, $\omega_\mathrm{*i}$ can then be tuned to be approximately constant in the region of interest. Figure~\ref{fig:modified_omega_star_profile} shows the profile of $\omega_\mathrm{*i}$ for the original and modified case. 
For this test, the adiabatic index $\Gamma$ is additionally set to zero. Then, the change in density is no longer related to the change in pressure and the continuity equation can be fully decoupled from the system of equations. Hence, we artificially keep the density constant in time and use the initial condition $\mathbf{v} = -\mathbf{v}_\mathrm{*i}$.  

\begin{figure}[htb]
    \centering
    \includegraphics[width=0.5\textwidth]{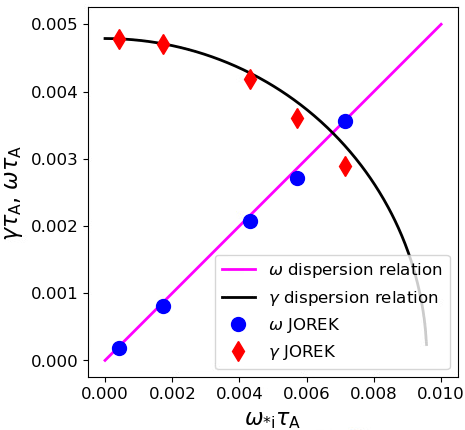}
    \caption{The mode frequency and growth rates obtained from JOREK simulations with diamagnetic terms for the modified input density and temperature profiles that keep $\omega_\mathrm{*i}$ approximately constant within the $q=1$ surface. The different data points were obtained by scanning the value of the central density $n_0$, which corresponds to a scan in the parameter $l_x$. The growth rate and frequency theoretically predicted by Equation~(\ref{eq:dispersion_relation}) are indicated by the solid lines.}
    \label{fig:quarter_circle}
\end{figure}

With the new setup, a scan in the density while keeping the pressure constant is done. Since $\omega_\mathrm{*i}$ is inversely proportional to the density and the ideal MHD growth rate for the kink mode without diamagnetic effects to its square root, the influence of $\omega_\mathrm{*i}$ can be analyzed in normalized units of $\tau_\mathrm{A} = \left(\mu_0 \rho_0\right)^{1/2} R_0 /B_\mathrm{ax} $, where $\rho_0$ is the mass density on axis. The results of this scan are presented in Figure~\ref{fig:quarter_circle}. The agreement between the analytical dispersion relation and the simulation results from JOREK is very good. In conclusion, it can be seen that the diamagnetic rotation stabilizes the internal kink mode. This effect becomes stronger for increasing $\omega_\mathrm{*i}$ (smaller density). The dispersion relation has also been reproduced by the CASTOR3D code for a different kink mode case \cite{Strumberger_2023}.

\subsubsection{Additional parameter scans.}

\begin{figure}[htb]
    \centering
    \includegraphics[width=0.5\textwidth]{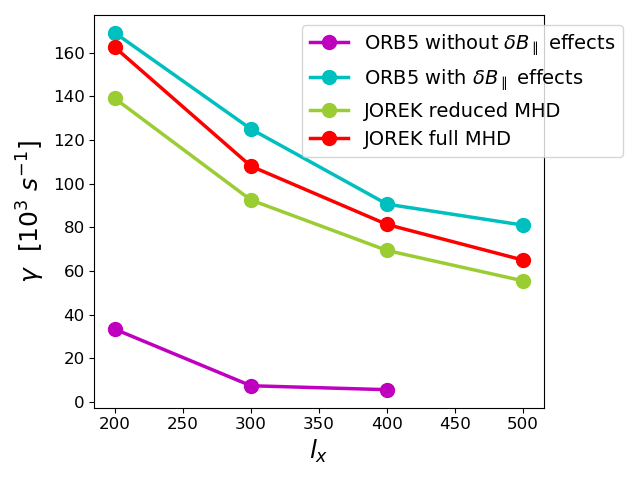}
    \caption{Comparison of the growth rate of the kink mode as a function of the parameter $l_x$ which controls the density to temperature ratio for a fixed value of $\beta_\mathrm{ORB5} = 0.0016$.}
    \label{fig:growth_rates_lx_scan}
\end{figure}

In the following, we consider again the original case with constant temperature and radially varying $\omega_\mathrm{*i}$ profile. 
A scan is done in the parameter $l_x$ for the single value of $\beta_\mathrm{ORB5} = 0.0016$. Changing $l_x$ while keeping the pressure constant effectively changes only a normalization in JOREK, i.e. the value of the particle density on axis. The latter scales $\propto l_x^2$. As the growth rate is proportional to $\tau_\mathrm{A}^{-1}$, and $\tau_\mathrm{A} \propto \rho_0^{1/2}$, the growth rate scales with $l_x^{-1}$. This is confirmed in the simulations, see Figure~\ref{fig:growth_rates_lx_scan}. The ORB5 simulations with $\delta B_\parallel$ effects show the same trend. In the case without $\delta B_\parallel$ effects, the growth rate is already very low such that the mode is nearly stabilized by increasing $l_x$.

\begin{figure}[htb]
    \centering
    \includegraphics[width=0.5\textwidth]{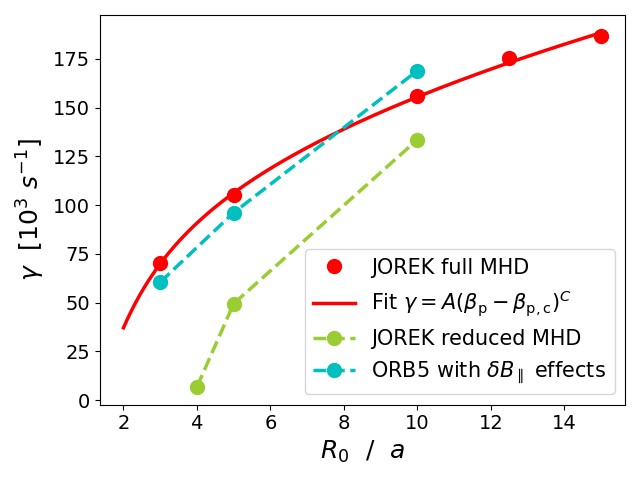}
    \caption{Comparison of the growth rates as a function of the aspect ratio $R_0/a$. 
    The red solid line represents a fit according to Equation (\ref{eq:kink_mode_growth_rate_fit}).}
    \label{fig:growth_rates_AR_scan}
\end{figure}

Another important parameter influencing the stability of the kink mode is the geometry under consideration. Therefore, a scan in the tokamak aspect ratio $R_0 / a$ is carried out by keeping $a=1\,\mathrm{m}$ and varying $R_0$. The results are presented in Figure~\ref{fig:growth_rates_AR_scan}. As pointed out in \cite{Bussac_1975}, the cylindrical approximation breaks down for $n=1$ modes such that the internal kink must be considered as a true toroidal instability. The growth rate derived in \cite{Bussac_1975} with the toroidal corrections takes the form
\begin{equation}
\label{eq:growth_rate_Bussac}
    \gamma \tau_\mathrm{A} = -\delta \hat{W} \sim \epsilon_1^2 \left(\beta_\mathrm{p}^2  - \beta_\mathrm{p,crit}^2\right),
\end{equation}
where $\delta \hat{W}$ is the normalized plasma potential energy, $\epsilon_1 = r_\mathrm{s} / R_0$ and $\beta_\mathrm{p,crit} $ can be estimated with Equation~(\ref{eq:beta_p,crit}) \cite{Porcelli_1996}. Since $\beta_\mathrm{p}$ scales with the poloidal magnetic field as $\beta_\mathrm{p} \propto B_m^{-2}(r_\mathrm{s})$ and $B_m(r_\mathrm{s}) \propto R_0^{-2}$ approximately, a stabilization of the kink mode is expected when the aspect ratio is decreased. This trend is also seen in Figure~\ref{fig:growth_rates_AR_scan}. In principle, a scan in the aspect ratio corresponds therefore also to a scan in $\beta_\mathrm{p}$ and a similar trend is found in comparison to Figure~\ref{fig:growth_rates_and_frequencies_beta_scan}.

The ORB5 and full MHD growth rates match well for all values of the aspect ratio. In reduced MHD, the kink mode is stabilized much stronger when reducing $R_0 / a$. This suggests that the assumptions made for the reduced MHD model break down for the kink mode at low values of the aspect ratio as the poloidal magnetic field becomes relatively larger compared to the toroidal field. As a consequence, the full 3D magnetic perturbations are needed to describe the kink mode accurately.

Previous numerical simulations showed that the growth rate of the ideal internal kink mode in realistic tokamak geometry scales rather as 
\begin{equation}
\label{eq:kink_mode_growth_rate_fit}
    \gamma = A \left( \beta_\mathrm{p} - \beta_\mathrm{p,c}\right)^C,
\end{equation}
with $C$ between 1 and 2 in contrast to Equation~(\ref{eq:growth_rate_Bussac}) \cite{Martynov_2005}. This model function is fitted to the JOREK full MHD growth rates and shows a better agreement with the data compared to Equation~(\ref{eq:growth_rate_Bussac}). $A$ is assumed to scale as $\sim R_0^{-3}$ and $\beta_\mathrm{p}$ is calculated explicitly for every data point according to Equation~(\ref{eq:beta_poloidal}). It scales approximately as $ \sim R_0^4$. We find $C\approx1.72$ and $\beta_\mathrm{p,c}\approx0.09$. This value also agrees reasonably well with $\beta_\mathrm{p,crit} \approx 0.12 $ obtained from Equation~(\ref{eq:beta_p,crit}). A possible reason for the discrepancy with Equation~(\ref{eq:growth_rate_Bussac}) for large $R_0 / a$ is the Shafranov shift that leads to a distortion of the circular flux surfaces on which the calculation in \cite{Bussac_1975} is based on. For instance, the Shafranov shift at $R_0 / a = 10$ is $\Delta = 0.18 \, \mathrm{m}$, which is almost 20\% of the minor radius. The shift increases at larger $R_0 / a$ as $\Delta $ increases with $\beta_\mathrm{p}$.

\begin{figure}[htb]
     \centering
     \begin{subfigure}[b]{0.49\textwidth}
         \centering
         \includegraphics[width=\textwidth]{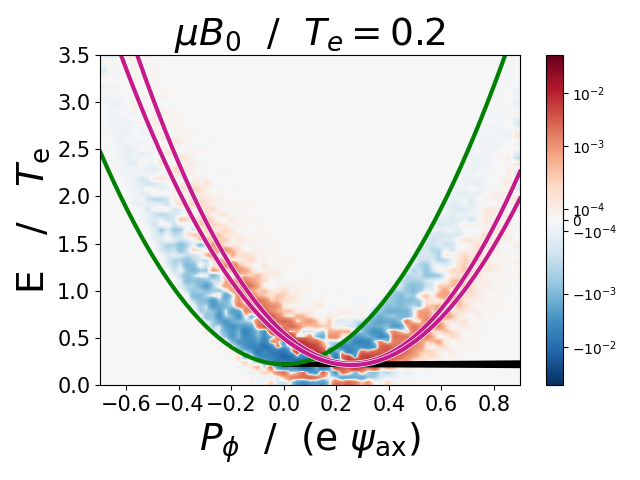}
         \caption{}
         \label{fig:pszs1}
     \end{subfigure}
     \hfill
     \begin{subfigure}[b]{0.49\textwidth}
         \centering
         \includegraphics[width=\textwidth]{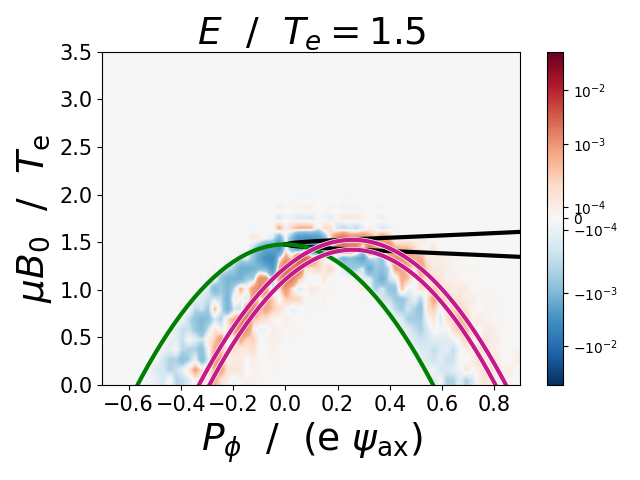}
         \caption{}
         \label{fig:pszs2}
     \end{subfigure}
     \begin{subfigure}[b]{0.49\textwidth}
         \centering
         \includegraphics[width=\textwidth]{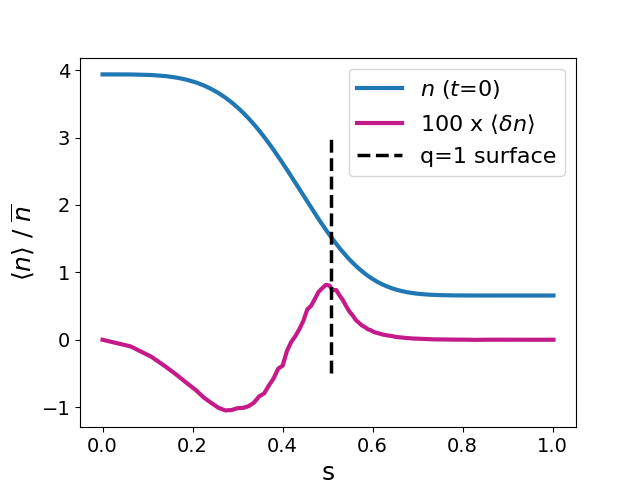}
         \caption{}
         \label{fig:pszs3}
     \end{subfigure}
        \caption{(a) Phase space zonal structure in energy $E$ and toroidal canonical momentum $P_\phi$ space for a fixed value of the magnetic moment $\mu$ and (b) in $\mu$-$P_\phi$ space for a fixed value of $E$. The trapped-passing particle boundary is marked by the black solid line. Orbits touching the magnetic axis are represented by the green solid line and orbits crossing the $q=1$ surface at the low field and high field side are denoted by the purple lines. (c) The flux surface average of the density perturbation $\left<\delta n\right>$ compared to the initial density profile $n(t=0)$ and normalized to the volume averaged density $\bar{n} $ as a function of the radial coordinate $s$. The perturbed density has its maximum approximately at the $q=1$ surface.}
        \label{fig:pszs_linear_phase}
\end{figure}

In order to understand in further detail the particle dynamics in connection with the kink mode, the phase space zonal structure (PSZS) diagnostic tool in ORB5 \cite{Bottino_2022} can be used to analyze the perturbation of the particle distribution function in the phase space. By averaging over angle variables, the phase space structures, that play an important role e.g. for the transport of the ions, are revealed. Figure~\ref{fig:pszs_linear_phase} shows the PSZS for the perturbed ion distribution function for the unstable kink base case in energy $E$ and canonical toroidal momentum $P_\phi$ space for fixed magnetic moment $\mu$ (see Figure~\ref{fig:pszs1}), and in $\mu$ - $P_\phi$ space for fixed $E$ (Figure~\ref{fig:pszs2}). Additional lines have been added which represent particle orbits touching the magnetic axis (green) and the $q$=1 surface on the high-field and low-field side (purple), and the trapped-passing boundary (black) \cite{White_2001}. The time point for this diagnostic is chosen in the later linear stage of the simulation. Since a large aspect ratio tokamak is considered, the fraction of trapped particles is very low. As can be seen from the figure, the region in phase space with positive change of the distribution function corresponds to the particle orbits near the $q$=1 surface. Figure~\ref{fig:pszs3} shows the flux surface average of the perturbed density as a function of the real space coordinate $s$ for the same time point. The flux surface average is computed by integrating over the toroidal and poloidal angles, too, such that that also only the zonal ($n=0, m=0$) component contributes. Analogously, $\left<\delta n\right>$ is positive and peaks approximately at the $q=1$ surface, becomes negative for smaller $s$ and is approximately zero at the magnetic axis. Unlike in the case of toroidal Alfv\'en eigenmodes (TAEs) for instance, the density profile does not flatten locally at the rational surface. The kink mode is rather a global mode in the sense that the total core region inside the $q=1$ surface is affected as for example, the radial displacement is nearly constant inside this region. The zonal density perturbation is already a nonlinear effect resulting from the coupling of the $m/n$ Fourier components and their negatives. There is no particular dependence of the zonal structures in velocity space; they are fully determined by the real space parameters.

\section{Conclusion and outlook}
\label{sec:conclusion}

In summary, in this work gyrokinetic and MHD simulations of $m/n = 1/1$ modes have been carried out with the ORB5 and JOREK numerical codes and compared to each other and to analytical theory. Supporting data was also provided by the CASTOR3D and LIGKA codes. It has been found that the ideal MHD internal kink mode can be indeed excited in a gyrokinetic simulation without collisions and using a shifted Maxwellian for the electron distribution function to account for the parallel current, which is part of the drive of the mode. Additionally, a finite pressure gradient is required for the mode to become unstable. The kink mode has been found to be very sensitive to the input parameters and how the simulation is exactly set up. One of these parameters is the electron mass, which is often increased artificially in gyrokinetic simulations to reduce the computational cost, but has been shown to have strong effects on the dynamics of the 1/1 mode. The collisionless tearing mode, which becomes strongly unstable if the electron-to-ion mass ratio is large and the plasma $\beta$ is low, is one example. This mode can not be found with MHD codes when they assume the electrons to be massless. However, it is important to note that there are extended MHD models which do include finite electron inertia and can capture collisionless tearing modes \cite{Guo_2020}.
Furthermore, it has been shown that it is important to take parallel magnetic field fluctuations $\delta B_\parallel$ into account for the internal kink mode. In ORB5, a simplified method is used for this purpose by just modifying the drift velocity $\mathbf{v}_\mathrm{d}$ according to Equation~(\ref{eq:modified_drift_velocity}) and not explicitly solving for $\delta B_\parallel$. However, taking into account this first order approximation has shown to have a strong effect on the mode structure, growth rate and mode frequency. A method for explicitly solving for the perturbed parallel magnetic field is currently being implemented in ORB5.
On the other hand, the reduced and full MHD models in JOREK led to very similar results, which may be due to the moderate value of the plasma $\beta$ in the considered cases. However, deviations between the reduced and full MHD model have been found at small aspect ratio like expected. 
The mode frequency has also been analyzed in the extended MHD model when enabling the diamagnetic terms and compared to the analytical dispersion relation for the internal kink mode. The latter with $\omega = \omega_\mathrm{*i} / 2$ can be reproduced in limiting cases when the ion diamagnetic frequency is nearly constant within the $q$=1 surface. In case of a radially varying $\omega_\mathrm{*i}$, the global nature of the mode leads to an averaging effect not captured by the analytical estimate.  
Furthermore, this global nature of the mode is also reflected in the fact that no local flattening of the density profile at the $q=1$ surface was observed in the late linear stage of the simulations. Moreover, no particular dependence of the zonal structures in the phase space was apparent.

As a next step, we plan to include also supra-thermal ions in the simulations for the studies of burning plasmas. This can be done in ORB5 as an additional particle species and in JOREK using the full-f kinetic particle extension for the fast ions \cite{Bogaarts_2022}. With this, we aim to compare the fishbone instability between the hybrid kinetic-MHD and the fully gyrokinetic approach. 

\section{Acknowledgments}
Discussions with S. Günter, A. Bottino, A. Biancalani and F. Widmer are gratefully acknowledged. This work has been carried out within the framework of the EUROfusion Consortium, funded by the
European Union via the Euratom Research and Training Programme (Grant Agreement No 101052200
— EUROfusion). Views and opinions expressed are however those of the author(s) only and do not
necessarily reflect those of the European Union or the European Commission. Neither the European
Union nor the European Commission can be held responsible for them.
Some of the simulations were done on the Hawk supercomputer hosted at HLRS and
the Leonardo supercomputer hosted at CINECA contributing to the TSVV projects 08 Integrated Modelling of Transient MHD Events and 10 Physics of Burning Plasmas.


\section*{References}
\bibliographystyle{iopart-num}
\bibliography{references}

\end{document}